\documentclass{article}

\PassOptionsToPackage{numbers}{natbib}

\usepackage[preprint]{neurips_2025}

\usepackage[utf8]{inputenc} 
\usepackage[T1]{fontenc}    
\usepackage{hyperref}       
\usepackage{url}            
\usepackage{booktabs}       
\usepackage{amsfonts}       
\usepackage{nicefrac}       
\usepackage{microtype}      
\usepackage{xcolor}         

\usepackage{acronym}
\usepackage{eqnarray}
\usepackage{arydshln}
\usepackage{booktabs}
\usepackage{hyperref}
\usepackage{graphicx}
\usepackage{algorithmic}
\usepackage{algorithm}
\usepackage{array}
\usepackage{amsmath,amsfonts}
\usepackage{wrapfig}
\usepackage{subcaption}

\acrodef{ssl}[SSL]{Self-Supervised Learning}
\acrodef{sl}[SL]{Supervised Learning}
\acrodef{ci}[CI]{Confidence Interval}
\acrodef{ast}[AST]{Audio Spectrogram Transformer}
\acrodef{mat}[MATPAC]{masked latent prediction an classification}
\acrodef{mcl}[MCL]{Multiple Choice Learning}
\acrodef{mlm}[MLM]{Masked Language Modeling}
\acrodef{ema}[EMA]{Exponential Moving Average}
\acrodef{mlp}[MaLaP]{Masked Latent Prediction}
\acrodef{met}[MATPAC++]{}
\acrodef{llm}[LLM]{Large Language Models}
\acrodef{alm}[ALM]{Audio-Language Models}
\acrodef{wta}[WTA]{Winner-Takes-All}

\newcommand{\pmval}[1]{{\fontsize{6pt}{6pt}\selectfont $\pm$ #1}}
\newcommand{\ie}{\textit{i.e.}}

\newcommand{\se}[1]{{\textcolor{black}{#1}}}
\newcommand{\secor}[2]{{\textcolor{black}{#2}}}
\newcommand{\secmt}[1]{}

\title{MATPAC++: Enhanced Masked Latent Prediction for Self-Supervised Audio Representation Learning}

\author{%
  Aurian Quelennec$^1$ \quad Pierre Chouteau$^1$ \quad Geoffroy Peeters$^1$ \quad Slim Essid$^1$\\ 
  $^1$LTCI, Télécom Paris, Institut Polytechnique de Paris\\
  \texttt{\{aurian.quelennec\}@telecom-paris.fr}\\
}

\begin{document}

\maketitle

\begin{abstract}
  Masked latent prediction has emerged as a leading paradigm in self-supervised learning (SSL), especially for general audio and music representation learning. While recent methods have demonstrated strong performance, the role of the predictor module used at the output of such SSL systems remains mainly overlooked, despite being crucial for solving the pretext task at hand.  In particular, this module should be able to deal with the ambiguity inherent in audio content, especially when it is composed of multiple sound sources.  This work proposes a novel enhancement: integrating Multiple Choice Learning (MCL) to explicitly model prediction ambiguity and improve representation quality. 
  We build on top of the recently proposed MATPAC system, improving its prediction and unsupervised classification pretext tasks with MCL.  We extensively evaluate our method, MATPAC++,  through both linear probing across multiple downstream tasks and fine-tuning on AudioSet, employing a unified protocol that enables rigorous and fair comparisons with state-of-the-art SSL approaches. Results show that our proposal achieves state-of-the-art when fine-tuned on AudioSet and overall state-of-the-art scores on downstream tasks. Additionally, we examine domain specialisation by training exclusively on music data, where our model achieves state-of-the-art performance with significantly improved efficiency. 
\end{abstract}

\section{Introduction}

With the rise of \ac{llm}, several multimodal \se{systems have exhibited promising capabilities} for solving vision or speech-related tasks \cite{Li23blip2, Alayrac22flamingo, moshi}. \secor{It is also the case for non-verbal speech sounds}{More recently, solutions supporting non-speech audio have been also developed}, where so-called \ac{alm} \cite{Wu23laion, elizalde23clap, gong24tlu, Gardner24llark, Ghosh24gamma, chu23qwen} have attempted to improve the perception and understanding of general audio and music by combining \ac{llm}s with powerful audio encoders.
Therefore, there is a strong interest in building robust audio encoders and clarifying how well they perform before they can be used in multimodal approaches to enhance audio understanding.  

A powerful approach to achieve such a goal is \se{Self-Supervised Learning} (\ac{ssl}), which has become a cornerstone in representation learning, \se{across various domains, including} general audio and music \cite{Liu22reviewaudio}, enabling models to learn rich and general representations without requiring ground-truth labels. 
Among all the existing pretext tasks that have been developed to train \ac{ssl} methods \cite{sslreview} \ac{mlp} has emerged as a leading paradigm, showing strong performance across different domains.  
Methods such as I-JEPA \cite{assran23ijepa}, M2D \cite{niizumi23m2d}, or MATPAC \cite{quelennec2025matpac} have demonstrated that learning to predict the latent representation of the masked parts of a partially visible input yields promising results.
In these methods, the predictor module plays a central role for the resolution of the \ac{mlp} pretext task.

\begin{wrapfigure}{r}{0.5\textwidth}
    \centering
    \includegraphics[width=\linewidth]{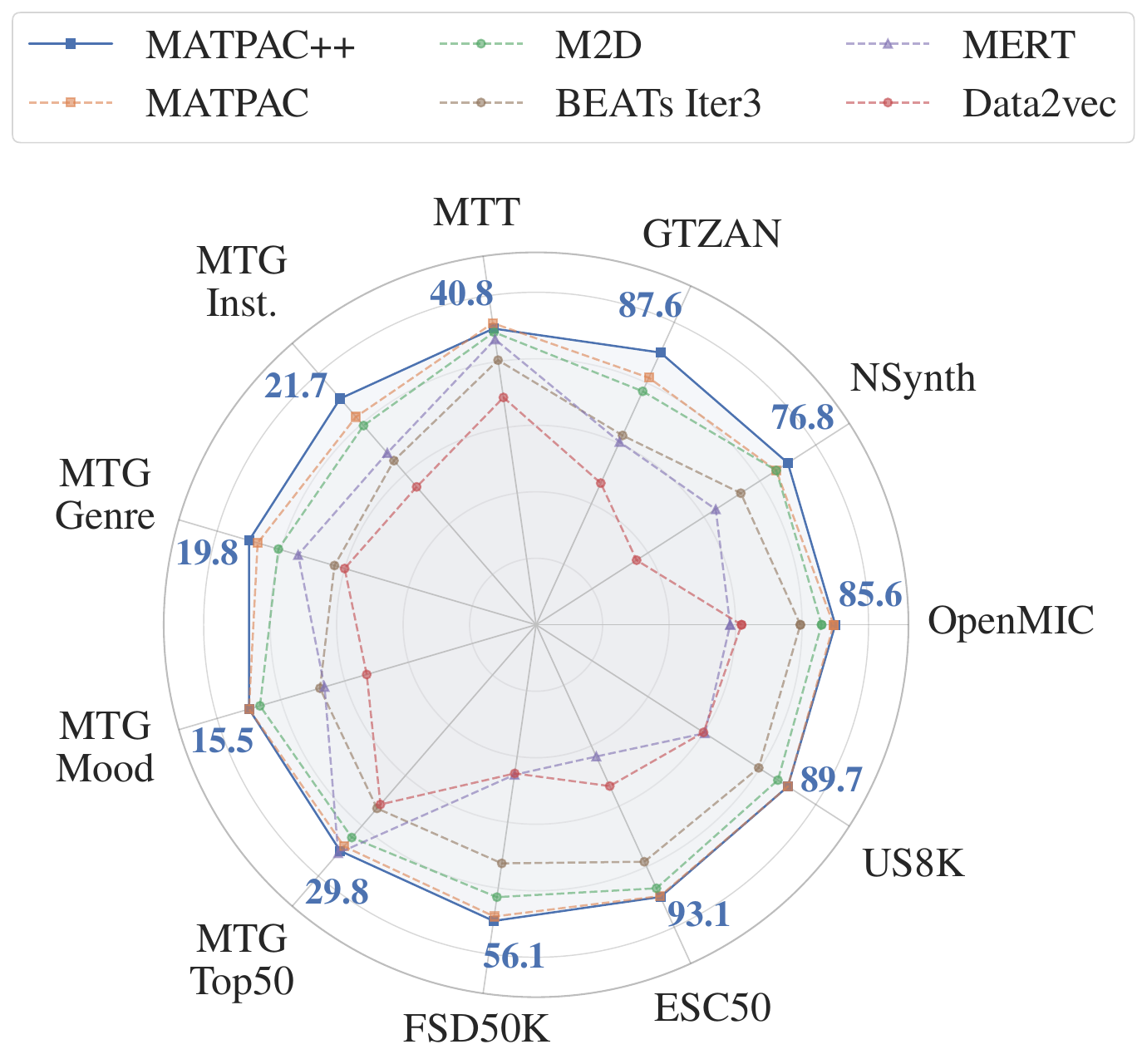}
    \vspace{-0.155cm}
    \caption{\acs{met} performances on reference datasets compared to \se{leading SSL methods:} MATPAC \cite{quelennec2025matpac}, M2D \cite{niizumi23m2d}, BEATs Iter3 \cite{chen23beats}, Data2vec \cite{baevski22data2vec} for general audio \ac{ssl} methods, MERT \cite{Li24mert} for music \ac{ssl} methods. 
    Best viewed in colour.}
    \vspace{-0.3cm}
    \label{fig:spider_graph}
\end{wrapfigure}

It is responsible for predicting the latent representations of the masked part of the input, encoded by a teacher model, only from the latent representations of the visible parts of the input, encoded by a student model.

However, predicting masked segments of an audio signal, based on visible parts, is inherently ambiguous:  there is no guarantee that an audio event will persist across patches, that its ``timbre'' will remain unchanged, or that new sound sources will not concurrently appear.
This makes the prediction task inherently one-to-many: multiple plausible latent completions may exist for a given input context, yet all existing \ac{mlp} methods output a single one.
We therefore propose here a novel enhancement to the \ac{mlp} framework in order to address this limitation, by introducing a \ac{mcl} component \cite{Lee16mcl, letzelter23mcl}. 
\ac{mcl} provides a simple way to take into account \secor{uncertainty}{ambiguity} by modelling multiple prediction hypotheses and selecting the most appropriate one, \se{using a variant of the so-called \ac{wta} strategy}. This naturally captures the diversity of plausible latent representations of the masked parts, which leads to more robust and semantically richer representations. 
We implement this in the state-of-the-art \ac{ssl} audio representation method \acs{mat} \cite{quelennec2025matpac}, resulting in a new model that we call \acs{met}.   

\textbf{Contributions.}
(i) We introduce \acs{met} a novel method to solve the ambiguity of the \ac{mlp} pretext task in self-supervised audio representation learning methods (multiple plausible latent completions may exist for a given input context). 
This is done by considering multiple output hypotheses in the predictor of the student branch and training using \ac{mcl}. 
(ii) We conduct an extensive experimental evaluation of our method and other state-of-the-art \ac{ssl} methods \secor{in both downstream tasks and fine-tuning on AudioSet}{on diverse tasks, including general audio and music classification tasks}. \se{In contrast to previous works, we dedicate a lot of care to the experimental validation, ensuring all models are compared in a fair manner, using exactly the same train/test data splits and careful examination of the statistical significance of the evaluation metrics considered.} 
We also perform a unified fine-tuning evaluation of all models using two distinct settings, revealing the limited reliability of previous evaluations. 
Overall, our method achieves \ac{ssl} state-of-the-art results on all tasks, as summarised in Fig. \ref{fig:spider_graph}. 
(iii) We examine domain specialisation by training the models exclusively on music data, again achieving state-of-the-art performance with a lighter model than previous ones. 

This work marks, to our knowledge, the first use of \ac{mcl} in general audio representation learning, and we provide insights into how it tackles ambiguity in the \ac{mlp} pretext task. Model checkpoints and code will be made publicly available. \footnote{\href{https://github.com/aurianworld/matpac}{https://github.com/aurianworld/matpac}}


\section{Related Works}
\label{RelatedWorks}

\textbf{Masked Latent Prediction pretext tasks.}
Among \ac{ssl} methods that rely on pretext tasks defined in a latent space, \acf{mlp} has been shown to be one of the most effective. In this case, a student encodes a masked part of the input and a predictor uses the student's latent representation to match the teacher's latent representation of either the full input or the complement of the masked input. 
Using this paradigm, M2D \cite{niizumi23m2d} for audio and I-JEPA \cite{assran23ijepa} for images, have achieved a strong performance while increasing the model's efficiency.
Consequently, we adopt the \ac{mlp} pretext task as one of the training objectives for our model, while addressing its limitation in handling prediction ambiguities.

\textbf{Unsupervised classification pretext tasks.}
Instead of matching latent representations, one can consider a classification pretext task. In \ac{ssl} this classification task is unsupervised, hence such an approach requires ``pseudo'' labels generated by the models themselves or by an external \ac{ssl} method.
Methods such as BEATs and HuBERT, for general audio and speech, exploit ``pseudo'' labels generated by clustering methods. 
They need, however, to be trained iteratively to achieve optimal performance, which is a limitation.
In contrast, methods developed in the vision domain, such as DINO \cite{Caron21dino}, pass different augmentations of an input image to the student and teacher networks, projects them into a probability distribution, and learns to match them through classification.
This approach proved to be quite effective. However, choosing the proper augmentations is always a tedious task.

\textbf{Multiple pretext tasks.}
To learn robust representations, \ac{ssl} methods may combine multiple pretext tasks, leveraging the complementary strengths that each task provides. ASiT \cite{Ahmed24Asit} and MAE-AST \cite{Baade22maeast}, for instance, jointly employ a reconstruction-based pretext task inspired by masked Audio-MAE, alongside a contrastive classification objective. 
 MATPAC \cite{quelennec2025matpac} introduces a cascaded approach, first performing \ac{mlp}, followed by unsupervised classification that aligns ``pseudo'' classes extracted from the predicted and target latent representations of masked patches. Building on this framework, our method incorporates \ac{mcl}, considering multiple plausible output hypotheses in the \ac{mlp} pretext task. It then adapts its outcomes, \ie the selected best hypotheses, to the unsupervised classification pretext task used in MATPAC.

\textbf{Multiple Choice Learning.}
\ac{mcl} has been considered in several machine learning tasks in recent years, leveraging multi-hypothesis neural networks \cite{guzman12mcl, Lee16mcl, rupprecht17mcl, garcia21mcl, letzelter23mcl}\secmt{garde quand même une réf de Victor !}. These works have shown that \se{predicting a single outcome is} suboptimal for \se{ambiguous} tasks, such as temporal tracking and forecasting for example. In the audio domain, \ac{mcl} has also been applied to the complex problem of multi-speaker source separation, yielding performance on par with state-of-the-art methods \cite{perera25mcl}. Building on these insights, we hypothesise that incorporating multiple prediction heads via an \ac{mcl} strategy can mitigate ambiguity in the \ac{mlp} pretext task, leading to more robust and informative representations.

Further details on related methods and their underlying principles are provided in Appendix \ref{app:relatedworks}.


\section{Method}

In audio, predicting masked patches from partial inputs is inherently ambiguous, as multiple plausible latent completions may exist due to variability in timbre, source continuity, or the emergence of new audio events. Existing \ac{mlp} methods typically predict a single outcome for each masked patch, overlooking this one-to-many nature. To address this, we enhance the \ac{mlp} framework with a \ac{mcl} component, enabling the model to generate and select among multiple hypotheses, thus capturing a richer and more robust representation of the latent space. 

\subsection{\secor{Masked Latent Prediction pretext task with multiple choice Learning}{Overview of \acs{met}}}
\label{sec:method_mlp}

\begin{figure}[t]
  \centering
  \includegraphics[width=0.9\linewidth]{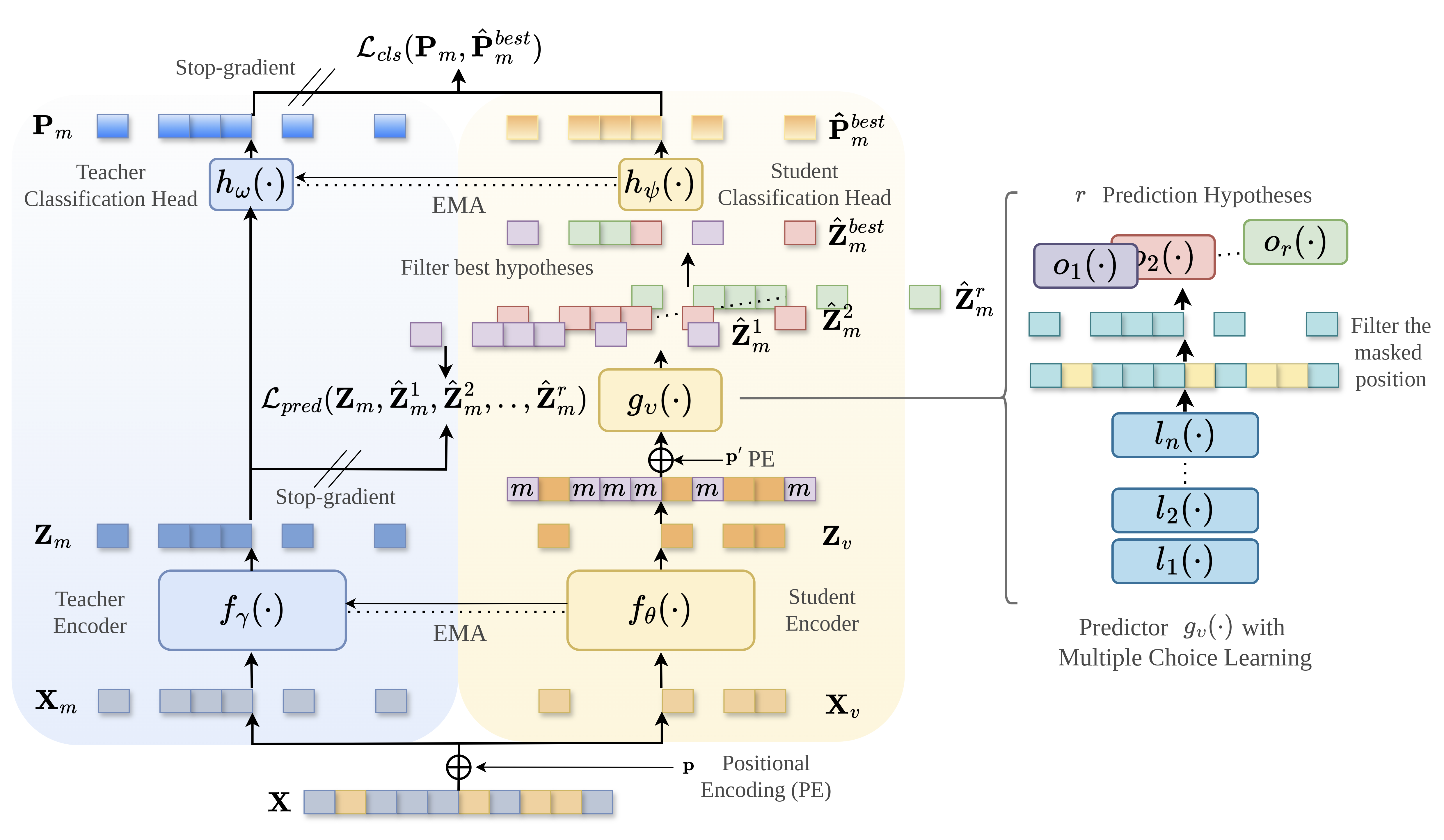}
  \caption{Block diagram of \acs{met} with details on the predictor's architecture.}
  \vspace{-0.3cm}
  \label{fig:flowchart}
\end{figure}

The block diagram of the model is given in Fig. \ref{fig:flowchart} and described below.

\textbf{Input of the model.} The input of the model is a set of 16 by 16 non-overlapping patches extracted from a log-scale Mel spectrogram. 
Each patch is then flattened and linearly projected into a 768-dimensional vector.
The resulting sequence for each audio sample is denoted by $\textbf{X}$.
A two-dimensional learnable positional encoding, $\textbf{p}$, is added to $\textbf{X}$, after which the sequence is randomly split into $\textbf{X}_v$, the visible patches, and $\textbf{X}_m$, the masked patches.
We adopt a random masking strategy with a masking ratio of 0.7.

\textbf{Projections in the latent space.} The student encoder $f_{\theta}$, maps the visible patch sequence $\textbf{X}_v$ to a latent representation $\textbf{Z}_v = f_{\theta}(\textbf{X}_v)$.
In parallel, the masked patches $\textbf{X}_m$ are processed by the teacher encoder $f{\gamma}$, yielding $\textbf{Z}_m = f_{\gamma}(\textbf{X}_m)$.
The teacher's parameters $\gamma$ are updated as an \ac{ema} of the student parameters $\theta$, following the rule $\gamma \xleftarrow{}\lambda\gamma + (1-\lambda)\theta$, where $\lambda$ denotes the decay rate.

\textbf{Multiple choice learning  prediction.} The predictor $g_{\upsilon}(.)$ takes as input the encoded visible patches, with a shared learnable token $\textit{\textbf{m}}$ appended at each masked position.
A new learnable positional embedding $\textbf{p}'$ is then added to provide location-specific information to the masked tokens. In standard \ac{mlp}-based approaches \cite{niizumi23m2d, assran23ijepa, quelennec2025matpac}, the predictor $g_{\upsilon}(.)$ consists of a sequence of transformer layers $\{l_1(.),l_2(.), \dots, l_n(.)\}_{n \in \mathbb{N}}$ followed by a single linear projection $o(.)$. The projection $o(.)$ maps the transformer layers' output to the encoders' latent space.
To incorporate \ac{mcl}, we introduce $r$ distinct linear projection heads ${o_1(.), o_2(.), \dots, o_r(.)}$, corresponding to $r$ different prediction hypotheses.
Given the latent representation $\textbf{Z}_v$ of the visible patches, our \ac{mcl} $g_{\upsilon}(.)$ produces $r$ separate hypotheses ${\hat{\textbf{Z}}^{(1)}_m, \hat{\textbf{Z}}^{(2)}_m, \dots, \hat{\textbf{Z}}^{(r)}_m}$ which are potential prediction of the $\textbf{Z}_m$ given by the teacher branch.
For each hypothesis $j \in \left[1, r\right]$, the $l_2$-normalized versions of $\textbf{Z}_m$ and each $\hat{\textbf{Z}}^{(j)}_m$, denoted as $\textbf{Z}'_m$ and $\hat{\textbf{Z}}'^{(j)}_m$, are compared using a mean square error: $\textbf{d}^{(j)}(i) = ||\hat{\textbf{z}}'^{(j)}_m(i) - \textbf{z}'_m(i) ||^2_2$ ;
with $i \in \left[1, N\right]$, and $N$ the number of targets, \textit{i.e.} the number of masked patches. 
We follow the annealed \ac{mcl} strategy introduced by Perera \cite{perera24mcl} to select the optimal hypothesis for each target, which mitigates convergence to suboptimal local minima.
The \ac{mcl} masked latent prediction loss is computed as a weighted sum based on the soft assignment of the best-performing hypothesis (which is specific to each patch $i$), \textit{i.e.}, the optimal $o_j(.)$, using
$\textbf{b}(i) = \text{Softmax}( -(  \textbf{d}^{(1)}(i), \cdots, \textbf{d}^{(r)}(i) )/ \tau_{mcl})$;
\begin{eqnarray}\label{eq:l_mcl_all}
    \mathcal{L}_{pred}(\textbf{Z}_m , \hat{\textbf{Z}}^{(1)}_m, \cdots, \hat{\textbf{Z}}^{(r)}_m) = \sum_{i \in \left[1,N\right]}{\sum_{j \in \left[1, r\right] }{\textbf{b}(i) \textbf{d}^{(j)}(i)}}; 
\end{eqnarray}
where $\tau_{mcl}$ is a temperature parameter that gradually decays during training, promoting exploration of the different hypotheses in the early stages.

\textbf{Multiple choice learning  classification.} Similarly to \cite{quelennec2025matpac}, we use an unsupervised classification pretext task on top of the \ac{mlp} pretext task. \se{This requires an adaptation to accommodate the \ac{mcl} component.}
While in \cite{quelennec2025matpac} the classification task is performed using the single prediction $\hat{\textbf{z}}_m(i)$ as input; in our \ac{mcl} case, we select for each target $i$, the best predictions among the $r$ possible ones, $\hat{\textbf{z}}^{(j)}_m(i)$, $j \in \left[1, r\right]$.
It is supposed to be the most informative one for the classification pretext task.
We perform this selection using the loss defined in Eq. \eqref{eq:l_mcl_all}.
For each masked patch $i$, we select the $\hat{\textbf{z}}^{(j)}_m(i)$ that minimizes the prediction loss: $\hat{\textbf{Z}}^{best}_{m} =  \{\textbf{z}_m'^{(k)}(i), k=\arg\min_{j \in \left[1,r\right]} \mathcal{L}^{(i,j)}_{pred}, i \in \left[1,N\right]\}$.

A student $h_{\psi}$ and teacher $h_{\omega}$ projection heads then transform the target and predicted latent representations into probability distributions of dimension $K$. 
Inspired by DINO \cite{Caron21dino} , we transform $\textbf{Z}_m$ and $\hat{\textbf{Z}}^{best}_m$ into the target distribution $\textbf{P}_m$ and the predicted distribution $\hat{\textbf{P}}^{best}_m$ using: 
$\hat{\textbf{P}}^{best}_m = \text{Softmax}( (h_{\psi}(\hat{\textbf{Z}}^{best}_m) / \tau_s)$;
$\textbf{P}_m = \text{Softmax}( (h_{\omega}(\textbf{Z}_m) - \textbf{C}) / \tau_t)$;
where $\tau_s$ and $\tau_t$ are temperature parameters used to sharpen the output distributions, while $\textbf{C}$ serves to center them.
$\textbf{C}$ is an \ac{ema} of the mean of $h_{\omega}(\textbf{Z}_m)$. 
The centering term $\textbf{C}$ is computed as an \ac{ema} of the mean of $h_{\omega}(\textbf{Z}_m)$.
These sharpening and centering techniques were originally introduced in DINO \cite{Caron21dino} to prevent model collapse and are key elements of MATPAC \cite{quelennec2025matpac} classification pretext task, where they serve the same purpose. Finally, the classification pretext task aligns the predicted and target distributions using a cross-entropy loss: 
\begin{equation}\label{eq:losscls}
    \mathcal{L}_{cls}(\hat{\textbf{P}}^{best}_m , \textbf{P}_m)= -\sum_{i=1}^{N}{\textbf{p}^{best}_m(i)\log(\hat{\textbf{p}}_m(i))}.
\end{equation}

The parameters $\omega$ of $h_{\omega}$ are updated as an \ac{ema} of $h_{\psi}$, using a decay rate $\zeta$ that is distinct from $\lambda$, the decay rate used in the \ac{ema} update of $f_{\theta}$.
\textbf{Final Loss.}
Our final training objective is the weighted sum of $\mathcal{L}_{cls}$ and $\mathcal{L}_{pred}$ using a parameter $\alpha$:
\begin{equation}\label{eq:final_loss}
   \mathcal{L} = (1-\alpha)\mathcal{L}_{cls} + \alpha \mathcal{L}_{mcl\ pred}.
\end{equation}


\section{Experiments}
We start by describing our rigorous experimental protocol allowing us to validate our proposal on diverse tasks and datasets, before discussing our results.

\subsection{Pre-training setup}

\textbf{Pre-training datasets.}
For pre-training, we randomly crop 6 seconds of content from each audio sample and transform it into a log-scale Mel spectrogram, we follow the processing used in \cite{niizumi23m2d, quelennec2025matpac}.
For general audio pre-training, we use AudioSet \cite{gemmeke2017audioset}, a dataset commonly used to pre-train \ac{ssl} methods, which enables a fair comparison with previous works. Our version contains over 2 million samples above 6-second duration. 
When pre-training on music-only data, we combine the Million Song Dataset \cite{Mahieux11msd} with \se{an in-house dataset music excerpts, covering 16 different music genres, which was captured from web radios, hereafter named} WebRadio. Together, the two datasets comprise a total of 1 million samples.
Please refer to Appendix \ref{app:pre_datasets} for more details.

\textbf{Model details.}
To ensure a rigorous comparison with prior work, we adopt the same encoder and predictor architectures as the ones used in \cite{niizumi23m2d, quelennec2025matpac}, based on Vision Transformers (ViT) \cite{dosovitskiy21vit}. 
For the classification heads, we follow the configuration of MATPAC \cite{quelennec2025matpac}, which itself builds upon the heads introduced in DINO \cite{Caron21dino}. 
All hyperparameters are retained as selected initially in these studies when pre-training on AudioSet and music data. 

\label{sec:ds_exp}

\subsection{Evaluation protocols}
 \label{sec:eval_protocols}

\textbf{\se{Downstream} Tasks.}
For downstream evaluation of models pre-trained on AudioSet, we adopt the same set of classification tasks as in \cite{quelennec2025matpac}, which has most of the tasks considered in \cite{niizumi23m2d}, covering both music and environmental sound domains.
\textbf{Instrument classification} is evaluated using NSynth \cite{EngelRRDNES17nsynth}, a dataset of synthetic musical notes, and OpenMIC \cite{Humphrey18openmic}, which features real-world music recordings.
For \textbf{genre classification}, we consider GTZAN \cite{TzanetakisC02gtzan} with the corrected labels proposed in \cite{Sturm13gtzancorr}.
We also include the MagnaTagATune \cite{magnatag} dataset for \textbf{music auto-tagging}.
In the environmental sound category, we evaluate on FSD50K \cite{FonsecaFPFS22fsd50k} a \textbf{sound event recognition} dataset, ESC-50 \cite{Piczak15esc50} as part of a \textbf{sound event classification} task, as well as UrbanSound8K \cite{SalamonJB14us8k}. 
To achieve a more detailed evaluation of the models pre-trained on music data, we add the tasks from the MTG-Jamendo dataset \cite{bogdanov2019mtg}, which consist of \textbf{multi-label genre}, \textbf{instrument}, \textbf{mood} and \textbf{50-most popular tags} classification tasks. This dataset has the advantage of providing numerous audio samples of professionally recorded music. 
Details of the tasks' descriptions and statistics are provided in Appendix \ref{app:ds_tasks}.

\textbf{Linear Probing Protocol.} We follow the downstream evaluation protocol of \cite{quelennec2025matpac}, which trains a linear head on top of the time-averaged (if needed) embeddings produced by the models to map them to the classes considered in each task. We use fixed training hyperparameters for all functions to ensure fair and consistent comparison across models. We only differ from the training protocol of \cite{quelennec2025matpac} by the way we do the early stopping on each task. Depending on the task, we either use the validation loss or the validation accuracy or mean average precision. All hyperparameters used in the downstream protocol are given in Appendix \ref{app:ds_protocol}.  

\textbf{\se{Encoder Fine-tuning on AudioSet.}}
\se{We also evaluate the \ac{ssl} methods when the encoders are fine-tuned in a supervised fashion on} the 2M AudioSet training samples with their labels, while evaluating on the reference test set provided by the dataset. We use the same fine-tuning settings as in \cite{niizumi24m2dx}, with a LARS optimiser, SpecAugment \cite{specaugment}, Cosine annealing \cite{cosine} and weighted sampling. 
PaSST \cite{Koutini22passt} introduced patchout to improve learning capabilities of  the \ac{ast} when trained in a supervised fashion. Moreover, the authors of \cite{niizumi24m2dx} showed that it was also beneficial to use patchout when fine-tuning pretrained \ac{ssl} methods. Since all the \ac{ssl} methods that we evaluate are based on \ac{ast} for their encoders, we fine-tune each of them using patchout with the same settings as in \cite{niizumi24m2dx}. 
In addition, to have a comparison point, we also fine-tune all \ac{ssl} methods without patchout, as regularly done, but with the same set of training hyperparameters for all models. This is to enable the best comparison possible. 
Finally, when publicly available, we evaluate the fine-tuned encoder and classification head of each \ac{ssl} model, \se{using the original authors' checkpoint}, directly on the test set of \textbf{our} AudioSet version. This setup offers a fair and informative assessment, given that variations in AudioSet versions, caused by unavailable or removed videos, make result comparison across studies quite challenging. 


\subsection{Results and discussion}

\begin{table}[h!]
    \footnotesize
    \caption{Downstream evaluation comparison with \ac{ci} at 95\% when trained on AudioSet. We greyed out supervised methods. Scores in bold are the best \ac{ssl} results and the ones underlined are the best scores overall.}
    \centering
    \resizebox{0.95\textwidth}{!}{
    \begin{tabular}{ l l l l l l l l l l}
             & OpenMIC & NSynth & GTZAN & MTT & FSD50K & ESC50 & US8K & Avg.\\
        Models & mAP & Acc(\%) & Acc(\%) & mAP & mAP & Acc(\%) & Acc(\%)\\
        \midrule

         \textcolor{gray}{HTS-AT}\textcolor{gray}{\cite{Chen22htsat}} & \textcolor{gray}{86.3\pmval{0.0}} & \textcolor{gray}{70.0\pmval{0.2}} & \textcolor{gray}{85.5\pmval{0.4}} & \textcolor{gray}{40.1\pmval{0.0}} & \textcolor{gray}{59.7\pmval{0.0}} & \textcolor{gray}{95.9\pmval{0.1}} & \textcolor{gray}{85.7\pmval{0.1}} & \textcolor{gray}{74.8} \\
        
        \textcolor{gray}{BEATs iter3+}\textcolor{gray}{\cite{chen23beats}} & \textcolor{gray}{\underline{86.8}\pmval{0.1}} & \textcolor{gray}{73.4\pmval{0.2}} & \textcolor{gray}{85.5\pmval{0.0}} & \textcolor{gray}{40.4\pmval{0.1}} & \textcolor{gray}{60.1\pmval{0.7}} & \textcolor{gray}{96.0\pmval{0.1}} & \textcolor{gray}{\underline{90.0}\pmval{0.1}} & \textcolor{gray}{76.1}\\

        \textcolor{gray}{PaSST} \textcolor{gray}{\cite{Koutini22passt}} &  \textcolor{gray}{86.4\pmval{0.2}} & \textcolor{gray}{73.2\pmval{0.2}} & \textcolor{gray}{87.5\pmval{0.7}} & \textcolor{gray}{40.4\pmval{0.2}} & \textcolor{gray}{\underline{62.3}\pmval{0.1}} & \textcolor{gray}{\underline{96.9}\pmval{0.1}} & \textcolor{gray}{89.6\pmval{0.3}}  & \textcolor{gray}{76.7}\\

        \midrule

        HuBERT \cite{Hsu21HUBERT} & 78.1\pmval{0.0} & 59.6\pmval{1.4} & 74.2\pmval{0.8} & 36.8\pmval{0.0} & 35.0\pmval{0.1} & 74.5\pmval{0.2} & 78.1\pmval{0.2} & 62.3 \\
        
        Data2vec \cite{baevski22data2vec} & 79.4\pmval{0.2} & 65.9\pmval{0.1} & 74.4\pmval{0.7} & 36.7\pmval{0.1} & 37.4\pmval{0.1} & 78.2\pmval{0.2} & 78.3\pmval{0.3} & 64.3 \\

        MAE-AST \cite{Baade22maeast} & 79.6\pmval{0.0} & 73.5\pmval{0.3} & 65.9\pmval{0.8} & 37.2\pmval{0.1} & 42.7\pmval{0.0} & 84.6\pmval{0.2} & 82.6\pmval{0.1} & 66.6 \\

        Dasheng Base \cite{dinkel2024dasheng} & 81.2\pmval{0.1} & 73.5\pmval{0.1} & 79.1\pmval{1.0} & 39.9\pmval{0.0} & 45.3\pmval{0.0} & 87.1\pmval{0.1} & 84.8\pmval{0.2} & 70.1\\

        ATST-Frame \cite{LiSL24atst} & 83.2\pmval{0.0} & 70.3\pmval{1.4} & 80.3\pmval{0.9} & 39.3\pmval{0.1} & 46.6\pmval{0.1} & 87.6\pmval{0.1} & 83.6\pmval{0.0} & 70.1 \\
        
        BEATs iter3 \cite{chen23beats} &  83.5\pmval{0.0} & 74.2\pmval{0.1} & 80.0\pmval{0.4} & 39.0\pmval{0.0} & 49.4\pmval{0.0} & 88.9\pmval{0.1} & 86.1\pmval{0.2} & 71.6 \\

        Dasheng 0.6B \cite{dinkel2024dasheng} & 82.0\pmval{0.0} & 76.5\pmval{0.3} & 83.3\pmval{0.7} & 39.6\pmval{0.1} & 48.2\pmval{0.0} & 88.2\pmval{0.2} & 85.8\pmval{0.2} & 71.9\\

        ATST-Clip \cite{LiSL24atst} & 84.4\pmval{0.1} & 71.7\pmval{0.5} & 79.5\pmval{1.8} & 39.1\pmval{0.1} & 50.9\pmval{0.0} & 91.4\pmval{0.1} & 86.2\pmval{0.1} & 71.9 \\    

        ASiT \cite{Ahmed24Asit} & 85.2\pmval{0.1} & \textbf{\underline{77.0}}\pmval{0.3} & 81.5\pmval{0.5} & 40.6\pmval{0.1} & 49.9\pmval{0.1} & 87.1\pmval{0.4} & 88.9\pmval{0.1} & 72.9 \\

        M2D \cite{niizumi23m2d} & 84.8\pmval{0.0} & 76.2\pmval{0.6} & 84.3\pmval{1.3} & 40.6\pmval{0.1} & 53.4\pmval{0.1} & 92.1\pmval{0.4} & 88.5\pmval{0.3} & 74.3\\
        
         \midrule
        
        MATPAC \cite{quelennec2025matpac} & 85.5\pmval{0.0}$^{\ast}$ & 76.2\pmval{0.3} & 85.5\pmval{0.6} & \textbf{\underline{41.1}}\pmval{0.0} & 55.6\pmval{0.0} & 93.0\pmval{0.0}$^{\ast}$ & \textbf{89.7}\pmval{0.3} & 75.2 \\

        \acs{met} \tiny{(\textit{ours})} & \textbf{85.6}\pmval{0.1} & 76.8\pmval{0.2} & \textbf{\underline{87.6}}\pmval{0.0} & 40.8\pmval{0.1} & \textbf{56.1}\pmval{0.1} & \textbf{93.1}\pmval{0.1} & \textbf{89.7}\pmval{0.3} & \textbf{75.7} \\

        \midrule
        
    \multicolumn{10}{l}{$^{\ast}$ $p < 0.05$, $^{\ast\ast}$ $p < 0.01$ compared to best model when CI are overlapping}      
    \end{tabular}
    }
    \label{tab:general}
\end{table}
\textbf{General Audio Pre-training.}
Table \ref{tab:general} presents the linear probing evaluation benchmark comparing \acs{met} with prior self-supervised and supervised models trained on AudioSet. \se{Multi-class classification tasks are evaluated using average accuracy, while multi-label tasks are evaluated through mean average precision (mAP).  All results are given along their $95\%$ confidence intervals (CI)}.
Our method achieves the highest average performance with a score of 75.7 \se{(average acroos all metrics)}, outperforming all other \ac{ssl} methods, and nearly closing the gap with the top-performing supervised baselines such as PaSST, which has an average score of 76.7. 
In particular, we set new SSL best scores on OpenMIC, FSD50K and ESC-50.
While \acs{met} outperforms supervised methods on NSynth and MTT, it outperforms supervised and self-supervised learning methods with an accuracy of 87.6\% on GTZAN.\secmt{Somethin's wrong in this previous statement!!}
Compared to MATPAC\footnote{The performance differences with \cite{quelennec2025matpac} stem from changes in the downstream evaluation protocol (Sec. \ref{sec:ds_exp}) and retraining on our version of AudioSet using the same hardware setup as \acs{met}. Notably, we report a higher average score than the original paper.} the improvements are consistent across five tasks: OpenMIC\secmt{You can't claim this without a test given the CI}, NSynth, GTZAN, FSD50K and ESC-50\secmt{same here}, and we have similar results on US8K. 
These results highlight the benefits of incorporating \ac{mcl} in the predictor's architecture. This design effectively addresses the inherent ambiguity of \ac{mlp} tasks and facilitates learning audio representations that generalise more robustly across diverse downstream settings.
This is particularly advantageous, as it introduces no additional complexity to the \secor{model's architecture or inference process}{the encoder used during inference}.

\textbf{Music Pre-training.}
To demonstrate the method's ability to specialise through domain-specific pre-training, we evaluate its performance on several music-focused downstream tasks, as shown in Table \ref{tab:music_res}. 
Since AudioSet contains many music-labelled samples, we also include results for \acs{met}, MATPAC, and M2D when pre-trained on AudioSet, serving as a general-audio baseline. Notably, \acs{met} pre-trained on AudioSet consistently matches or outperforms specialised music SSL models such as MERT \cite{Li24mert} and MULE \cite{mcallum22mule}. Furthermore, when pre-trained on music data, \acs{met} achieves the best performance on OpenMIC, MTT, and MTG Genre, and sets a new state-of-the-art on NSynth with an accuracy of 79\%.
As anticipated, pre-training on music-specific data improves \acs{met}'s performance on music tasks, raising its average score from 47.2 to 47.9. 
Interestingly, this gain is more modest than that observed for MATPAC, which improves from 46.8 to 47.9 under the same conditions. This narrower gap suggests that using \ac{mcl} when pre-training on a dataset with a large diversity and inherent ambiguity fosters the learning of more generalizable representations, but is less efficient when dealing with more structured data such as professional music recordings.
Nonetheless, these results demonstrate that \acs{met} can effectively leverage general and domain-specific data, offering a state-of-the-art \ac{ssl} solution for music understanding. 
Moreover, \acs{met} achieves this performance with significantly fewer parameters, 86M compared to MERT’s 330M, highlighting its efficiency in both representation learning and model capacity.
\begin{table}[h!]
    \footnotesize
    \caption{Downstream evaluation comparison with \ac{ci} at 95\%  on music tasks. \textit{Training Data} report the type of audio used for pre-training.}
    \centering
    \resizebox{\textwidth}{!}{
    \begin{tabular}{ l l l l l l l l l l l}
            & Training & OpenMIC & NSynth & GTZAN & MTT & MTG$_{\textit{Inst.}}$ & MTG$_{\textit{Genre.}}$ & MTG$_{\textit{Mood}}$ & MTG$_{\textit{Top50}}$ & Avg. \\
        Models &  Data & mAP & Acc(\%) & Acc(\%) & mAP & mAP & mAP & mAP & mAP & \\
        \midrule

        MULE$^{\dagger}$ \cite{mcallum22mule} & Music & - & 74.0 & 73.5 & 40.4 & 19.2 & 20.4 & 15.4 & \textbf{30.6} & - \\
        
        MERT$^{\dagger}$ \cite{Li24mert} & Music & 78.6\pmval{0.0} & 72.6 & 79.3 & 40.2 & 19.8 & 18.6 & 14.0 & 29.9 & 44.2 \\   

        M2D \cite{niizumi23m2d} & General & 84.8\pmval{0.0} & 76.2\pmval{0.6} & 84.3\pmval{1.3} & 40.6\pmval{0.1} & 20.8\pmval{0.1} & 19.1\pmval{0.1} & 15.3\pmval{0.3} & 29.2\pmval{0.0} & 46.3 \\

        \midrule 
        
        MATPAC \cite{quelennec2025matpac} & General & 85.5\pmval{0.0} & 76.2\pmval{0.3} & 85.5\pmval{0.6} & 41.1\pmval{0.0}$^{\ast}$ & 21.1\pmval{0.0} & 19.6\pmval{0.1} & 15.5\pmval{0.0} & 29.6\pmval{0.0} & 46.8 \\        

        MATPAC \cite{quelennec2025matpac} & Music & 85.8\pmval{0.0} & 77.7\pmval{0.2} & \textbf{88.8}\pmval{0.3} & 41.1\pmval{0.0}$^{\ast}$ & \textbf{22.8}\pmval{0.1} & \textbf{20.6}\pmval{0.0} & \textbf{15.7}\pmval{0.1} & 30.3\pmval{0.0} & \textbf{47.9} \\      
        
        \acs{met} \tiny{(\textit{ours})} & General & 85.6\pmval{0.1} & 76.8\pmval{0.2} & 87.6\pmval{0.0} & 40.8\pmval{0.1} & 21.7\pmval{0.0} & 19.8\pmval{0.0} & 15.5\pmval{0.1} & 29.8\pmval{0.0} & 47.2 \\
        
        \acs{met} \tiny{(\textit{ours})} & Music & \textbf{86.1}\pmval{0.0} & \textbf{79.0}\pmval{0.1} & 88.0\pmval{0.3} & \textbf{41.2}\pmval{0.1} & 22.4\pmval{0.0} & \textbf{20.6}\pmval{0.0} & 15.4\pmval{0.0} & 30.3\pmval{0.0} & \textbf{47.9} \\

        \midrule  
    \multicolumn{10}{l}{$^{\dagger}$ All the results come from \cite{Li24mert} except for OpenMIC.} $^{\ast}$ $p < 0.05$, $^{\ast\ast}$ $p < 0.01$ compared to best model when CI are overlapping
    \end{tabular}
    }
    \label{tab:music_res}
    \vspace{-0.3cm}
\end{table}

\textbf{Encoder Fine-tuning Evaluation.}
In addition to evaluating our method on downstream tasks, we also conduct a fine-tuning evaluation to assess its transferability. Fine-tuning on large annotated datasets such as AudioSet is complex and time-consuming, with numerous possible pipelines available \cite{Lora}. Due to this complexity, many studies rely on reported results from the literature rather than performing their own fine-tuning of other methods.
Moreover, a significant limitation of AudioSet is that it provides only links to YouTube videos for data retrieval. As a result, some videos may become unavailable or modified over time, making it difficult to ensure consistency across different versions of the dataset used in comparative studies.
For all these reasons, we decided to fine-tune all the reference \ac{ssl} audio methods in a unified framework on the same version of AudioSet we have access to. In Table \ref{tab:ft_results}, for each method, we report the results from the original paper, the evaluation of the fine-tuned model, \se{using the original authors' checkpoint} if publicly available, on our version of the test set of AudioSet, and finally our unified fine-tuning results under two different settings. \se{As can be seen in this table, dataset version differences may cause significant differences between previously reported results (column \textit{'Litt.'}) and the ones we reproduce, hence the latter is a fairer reference for comparison.}
\begin{wraptable}{r}{0.5\textwidth}
    \footnotesize
    \caption{Unified fine-tuning results on AudioSet.
    }
    \centering
    \resizebox{0.5\textwidth}{!}{
    \begin{tabular}{ l l l l l}
             & Litt. & Eval & PO FT & FT \\
        Models & mAP & mAP & mAP & mAP \\
        \midrule

         \textcolor{gray}{HTS-AT} \textcolor{gray}{ \cite{Chen22htsat}}& \textcolor{gray}{47.1 } & \textcolor{gray}{46.25} & \textcolor{gray}{-} & \textcolor{gray}{-}  \\
        
        \textcolor{gray}{BEATs iter3+}  \textcolor{gray}{ \cite{chen23beats}}& \textcolor{gray}{48.6 } & \textcolor{gray}{45.45} & \textcolor{gray}{-} & \textcolor{gray}{-} \\

        \textcolor{gray}{PaSST} \textcolor{gray}{ \cite{Koutini22passt}}& \textcolor{gray}{47.1 } & \textcolor{gray}{45.13} & \textcolor{gray}{-} & \textcolor{gray}{-} \\

        \midrule

        MAE-AST \cite{Baade22maeast} & - & - & 42.13$^{\ast\ast}$ & 43.05$^{\ast\ast}$ \\

        Dasheng Base \cite{dinkel2024dasheng} & - & - & 44.32$^{\ast\ast}$ & 44.65$^{\ast\ast}$ \\

        ATST-Frame \cite{LiSL24atst} & 48.0  & - & 38.02$^{\ast\ast}$ & 39.80$^{\ast\ast}$ \\
        
        BEATs iter3 \cite{chen23beats}  & 48.0  & 41.91 & 45.55$^{\ast\ast}$ & \textbf{46.98} \\

        Dasheng 0.6B \cite{dinkel2024dasheng} & - & - &  44.88$^{\ast\ast}$ & - \\

        ATST-Clip \cite{LiSL24atst} & 45.2  & - & 43.81$^{\ast\ast}$ & 44.17$^{\ast\ast}$ \\    

        ASiT \cite{Ahmed24Asit} & 47.5  & - & 45.26$^{\ast\ast}$ & 46.21$^{\ast\ast}$ \\

        M2D \cite{niizumi23m2d} & 47.9  & 47.83 & 47.89$^{\ast\ast}$ & 44.71$^{\ast\ast}$\\

        \midrule
        
        MATPAC \cite{quelennec2025matpac} & - & - & 47.96$^{\ast}$ & 46.85$^{\ast\ast}$ \\        
        
        \acs{met} \tiny{(\textit{ours})} & - & - & \textbf{\underline{48.09}} & 46.50$^{\ast\ast}$ \\
        
         \midrule
         \multicolumn{5}{l}{$^{\ast}$ $p < 0.05$, $^{\ast\ast}$ $p < 0.01$ compared to best model}\\
    \end{tabular}
    }
    \label{tab:ft_results}
    \vspace{-0.3cm}
\end{wraptable}

\textbf{Statistical Significance.}
\secmt{say this in the beginning and use it in the discussion}By fine-tuning all methods within a unified framework, we were able to apply a non-parametric Wilcoxon signed-rank test to assess statistical differences in model predictions on the AudioSet test set. Under patchout fine-tuning, the results indicate a marginal significant difference between the predictions of \acs{met} and MATPAC. For standard fine-tuning however, the score is statistically different between BEATS and the other methods, including \acs{met}. These statistical findings further underscore the challenges of comparing methods under fine-tuning on AudioSet, as methods' performance is highly sensitive to training configurations. One could design method-specific tuning routines to optimise results, potentially biasing comparisons.

\textbf{Patchout Fine-tuning.}
Since all considered \ac{ssl} methods employ \ac{ast} encoders, we follow the fine-tuning protocol proposed in \cite{niizumi24m2dx}, applying patchout, introduced in PaSST \cite{Koutini22passt}, which involves masking a portion of the transformer input sequence and performing classification based on the mean of the projected visible tokens. 
Interestingly, models that use \ac{mlp} as a pretext task show greater gains from patchout training. 
Under this setup, \acs{met} achieves a state-of-the-art score of 48.09, outperforming all evaluated \ac{ssl} methods in our unified fine-tuning benchmark and previously reported results in the literature.

\textbf{Standard Fine-tuning.}
To highlight the sensitivity of model performance to the fine-tuning strategy, we also report results under a standard fine-tuning protocol, where the full audio input is used for supervised classification on AudioSet.\footnote{Dascheng 0.6B could not finish training after 4 days of fine-tuning.} Under this setting, the ranking of methods differs from that observed with patchout fine-tuning: BEATs achieves the highest score with a mAP of 46.98\secmt{no, if your stars are correct BEATs and MATPAC are the same; say it like that}, followed by MATPAC and \acs{met} as the second and third best-performing models, respectively. 

Overall, we highlighted that \acs{met} achieves top performances compared to other \ac{ssl} methods when used in a transfer learning context. We also give more insights with bootstrap results in Appendix \ref{app:ft_results}.

\subsection{Ablation Studies}
\label{sec:ablation}

All ablation studies are conducted using a version of the method pre-trained on 3-second audio segments to reduce computational costs. In Appendix \ref{app:other_ablations}, we compare models trained on 3s and 6s segments and observe strongly correlated results, confirming the validity of using 3s segments. 

\begin{table}[h!]
    \footnotesize
    \caption{\ac{mcl} strategies for the selection of hypotheses.}
    \centering
    
    \resizebox{0.85\textwidth}{!}{
    \begin{tabular}{ l l l l l l l l l l }
    
              & OpenMIC & NSynth & GTZAN & MTT & FSD50K & ESC50 & US8K & Avg.\\
        Strategy & mAP & Acc(\%) & Acc(\%) & mAP & mAP & Acc(\%) & Acc(\%) & \\
         \midrule        

         Annealed & \textbf{85.6}\pmval{0.0} & 77.1\pmval{0.2} & \textbf{86.1}\pmval{0.3} & 40.9\pmval{0.0} & \textbf{55.8}\pmval{0.0} & 92.8\pmval{0.1} & \textbf{90.3}\pmval{0.1} & \textbf{75.5} \\
         
         \addlinespace[0.05cm] \hdashline \addlinespace[0.07cm]
         
         Mean & 85.2\pmval{0.0} & \textbf{79.6}\pmval{0.3} & 84.9\pmval{03} & 40.9\pmval{0.1} & 55.1\pmval{0.0} & 92.9\pmval{0.0} & 88.7\pmval{0.1} & 75.3 \\

         Greedy & 85.2\pmval{0.0} & 76.3\pmval{0.4} & 85.4\pmval{0.5} & 40.9\pmval{0.0} & 55.0\pmval{0.1} & \textbf{93.3}\pmval{0.3} & 89.1\pmval{0.0} & 75.0\\
                  
    \vspace{-0.3cm}
\end{tabular}
}
    \label{tab:mcl_method}
\end{table}
\begin{wrapfigure}{r}{0.5\textwidth}
    \vspace{-0.3cm}
    \centering
    \includegraphics[width=\linewidth]{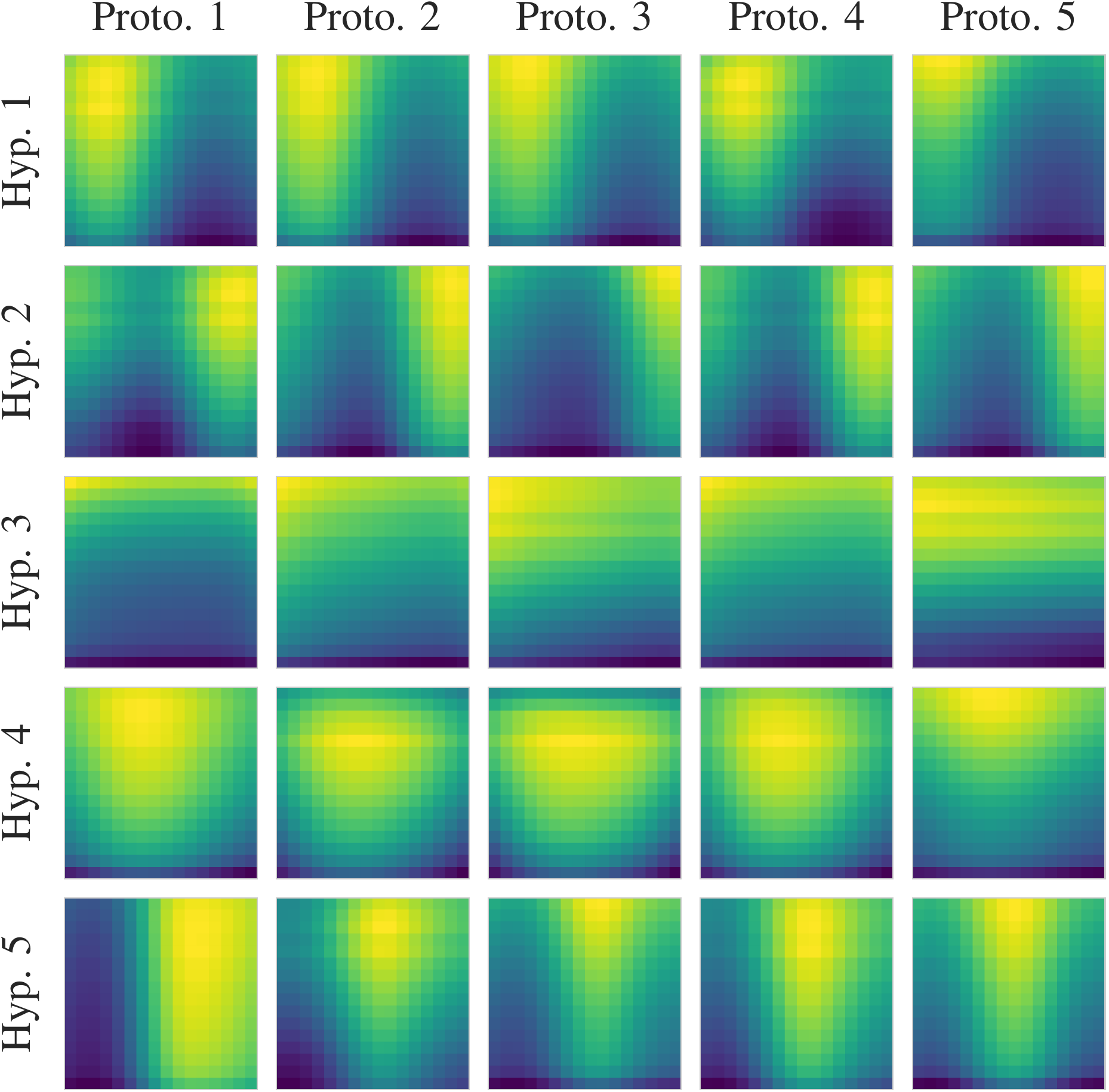}
    \vspace{-0.3cm}
    \caption{Cluster Prototypes for each Hypothesis. $x$ axis is the time, $y$ axis the frequencies.}
    \label{fig:prototypes}
    \vspace{-0.3cm}
\end{wrapfigure}

\textbf{Hypothesis selection strategies.}
Here, we investigate various hypothesis selection strategies within our multi-head output framework. We evaluate the mean of all hypotheses to demonstrate that the performance gains are not merely due to the increased parameter count from multiple heads. Next, we consider a greedy strategy, which selects the hypothesis minimising the prediction loss for each instance. Finally, we assess the annealed selection strategy \cite{perera24mcl}, used by default in \acs{met}, which balances exploration and exploitation during training.

Results in Table~\ref{tab:mcl_method} show that the annealed selection strategy consistently outperforms more naive approaches. However, on simpler tasks such as NSynth, comprising isolated instrument notes at fixed pitches, averaging the hypotheses yields a notable performance boost. While selecting among multiple hypotheses addresses ambiguity by promoting diverse and plausible outputs, averaging them instead enforces stability and consistency. This behaviour, \se{akin to a simplified form of ensembling,} appears better suited to low-variability audio, such as the synthetic signals found in NSynth.

\textbf{Hypothesis specialisation.}
We analyse the prediction behaviour of each head $o_i$ to examine whether each hypothesis specialises in distinct regions of the teacher encoder's latent space.
For 200,000 training samples\secmt{drawn randomly?}, we track the \secor{best-selected}{winning} hypothesis for each masked patch. 
We then apply K-means clustering (with $k=5$) on the set of all log-scale Mel spectrogram patches associated with each hypothesis and visualise the resulting cluster centroids, seen as hypothesis ``prototypes'', in Fig. \ref{fig:prototypes}. 
The results confirm that hypotheses specialise in complementary spectro-temporal patterns. 
For example, Hypothesis 1 focuses on energy concentrated on the left side of the patch, while Hypothesis 2 captures patterns on the right. 
Hypothesis 3 represents patches with uniform energy along time, whereas Hypothesis 5 captures vertical, frequency-focused energy patterns. 
These diverse specialisations validate the usefulness of our MCL component to promote a diverse set of plausible predictions for each masked time-frequency patch.\secmt{J'ai volontairement nuance un peu}

\begin{figure}
    \centering
    \includegraphics[width=0.9\linewidth]{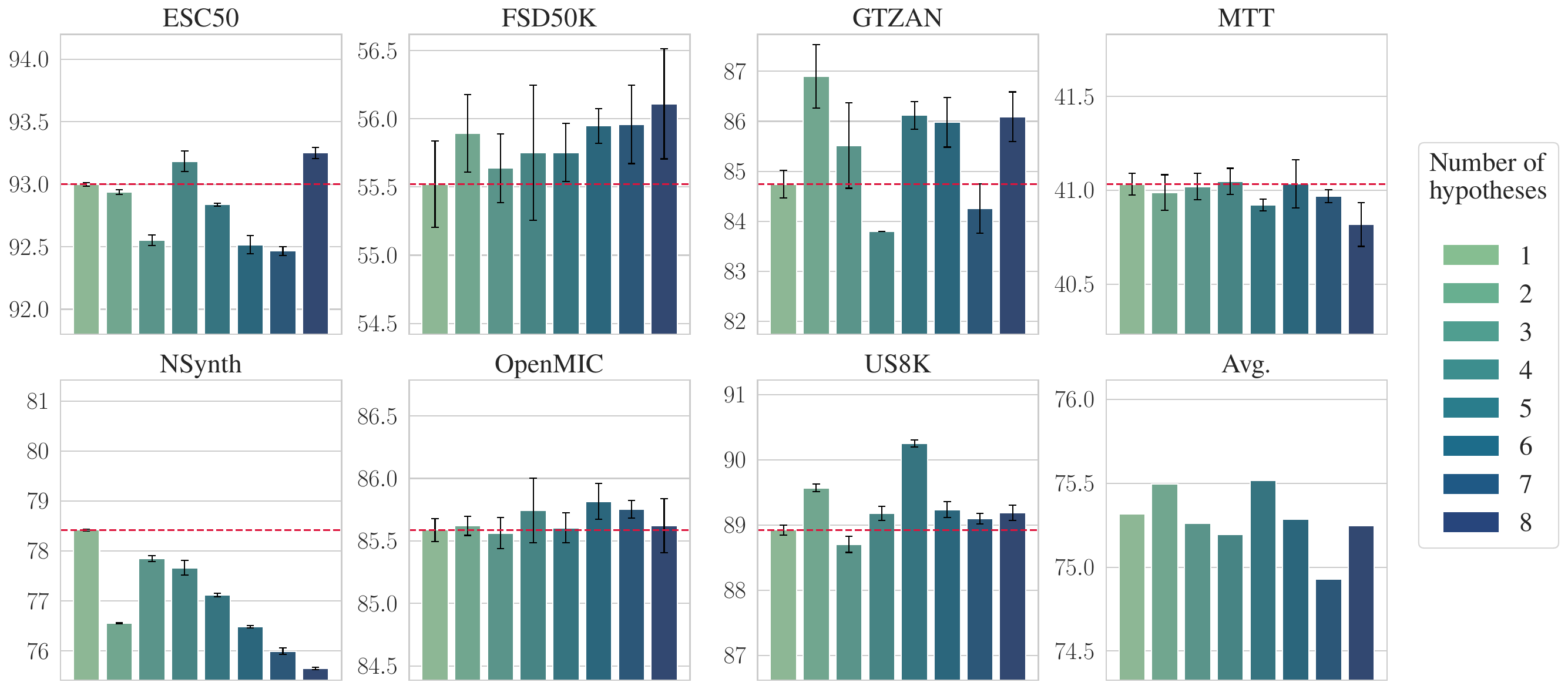}
    \caption{Impact of the number of hypotheses, models trained on 3s of audio.}
    \label{fig:head_ablation}
    \vspace{-0.3cm}
\end{figure}

\textbf{Number of hypotheses.}
We further conduct an ablation study to examine the impact of the number of hypotheses used in the \ac{mcl} component. Task-specific results are presented in Fig. \ref{fig:head_ablation}. Overall, incorporating multiple hypotheses generally improves performance over the single-head baseline. However, we observe that increasing the number of hypotheses degrades performance on simpler, low-variability tasks such as ESC-50 and NSynth, where consistency is more beneficial, as discussed in the previous ablation. In contrast, performance improves with more hypotheses on complex, multilabel tasks involving real-world audio, such as FSD50K, suggesting that diversity is more valuable in ambiguous or noisy settings, \se{as it is generally the case in real-world settings}.

Additional analyses and interpretability results for \acs{met} are provided in Appendix~\ref{app:other_ablations}.


\section{Conclusions}
In this work, we introduced \acs{met}, a novel self-supervised learning method that addresses the inherent ambiguity of the masked latent prediction  pretext-task by considering multiple output hypotheses in the predictor design and training using the multiple choice learning paradigm. Our extensive empirical evaluation, covering general audio and music classification tasks, demonstrates that \acs{met} consistently outperforms prior self-supervised methods across tasks, and its encoder is among the top-performing \ac{ssl} methods when fine-tuned. 
By ensuring a unified and statistically grounded comparison protocol, we reveal limitations in existing evaluation practices of encoder fine-tuning. Furthermore, MATPAC++ excels in domain-specialised training, achieving state-of-the-art performance on music tasks with significantly fewer parameters than existing models. 
This is the first application of \ac{mcl} for \ac{ssl} audio representation learning, offering new insights into how ambiguity can be modelled to produce richer, more generalizable audio representations.

\paragraph{Limitations}
One limitation of our approach is that each additional hypothesis increases the number of parameters in the predictor, thereby raising the model’s complexity during training (but not during inference).
Furthermore, although our results clearly show that multiple hypotheses help mitigate task ambiguity, the interpretability of what each hypothesis captures remains limited. Future work could focus on improving the reusability of the trained predictor, individual hypotheses, and classification heads at inference.


\newpage
\bibliographystyle{unsrtnat}
\bibliography{biblio}
\newpage


\appendix

The appendix is organised as follows: the first section presents additional experiments and results that provide further insights into our method's behaviour. This is followed by a detailed description of our experiments' training parameters, evaluation protocol, and datasets. Finally, we provide an extended and comprehensive literature review of related work.


\section{Supplementary Results}
\label{app:more-results}

\subsection{Encoder Fine-tuning Evaluation}
\label{app:ft_results}

\begin{figure}[h]
    \centering
    \includegraphics[width=0.8\linewidth]{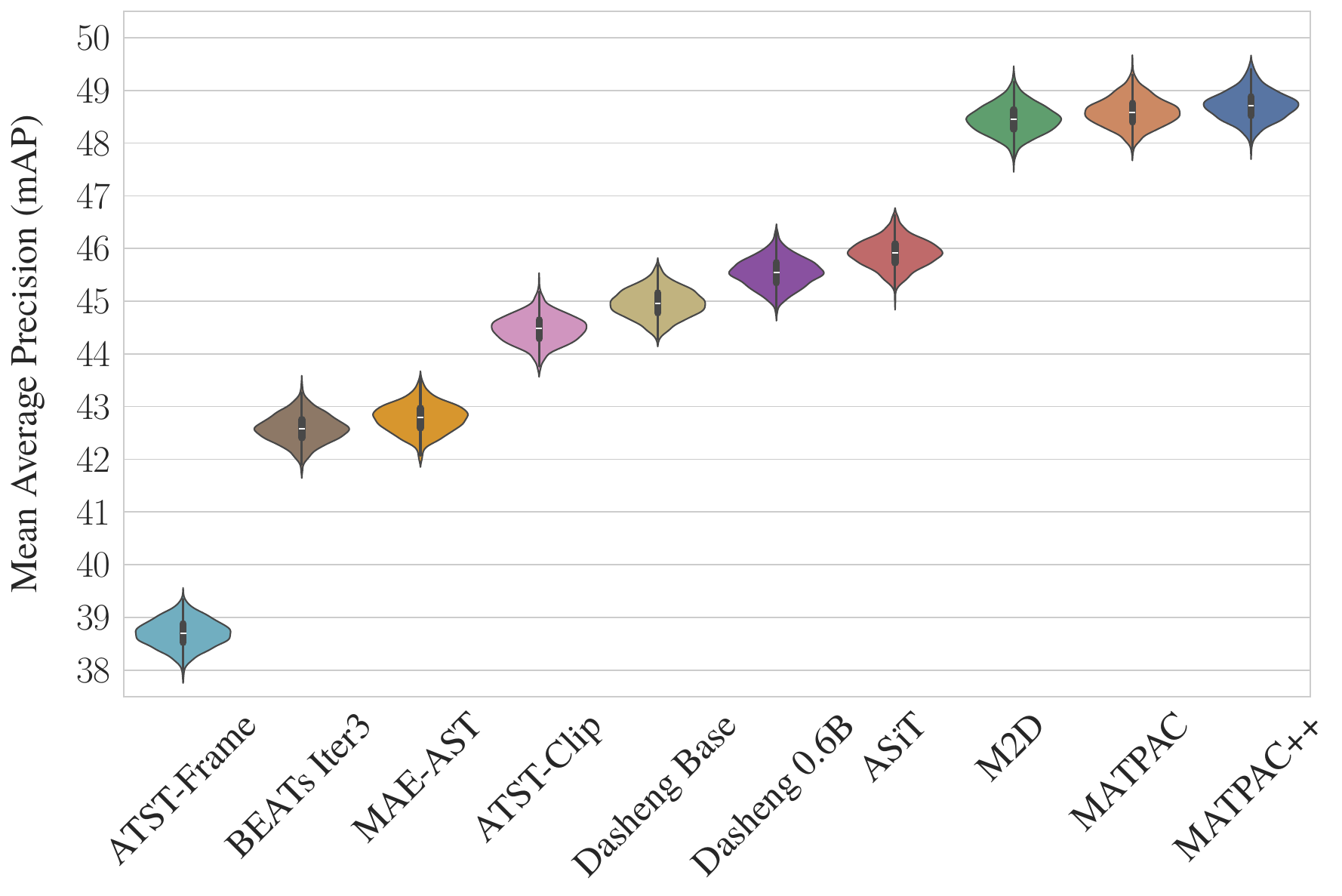}
    \caption{AudioSet bootstrap evaluation for patchout fine-tuning.}
    \label{fig:as_bootstrap_po_ft}
\end{figure}

\begin{figure}[h]
    \centering
    \includegraphics[width=0.8\linewidth]{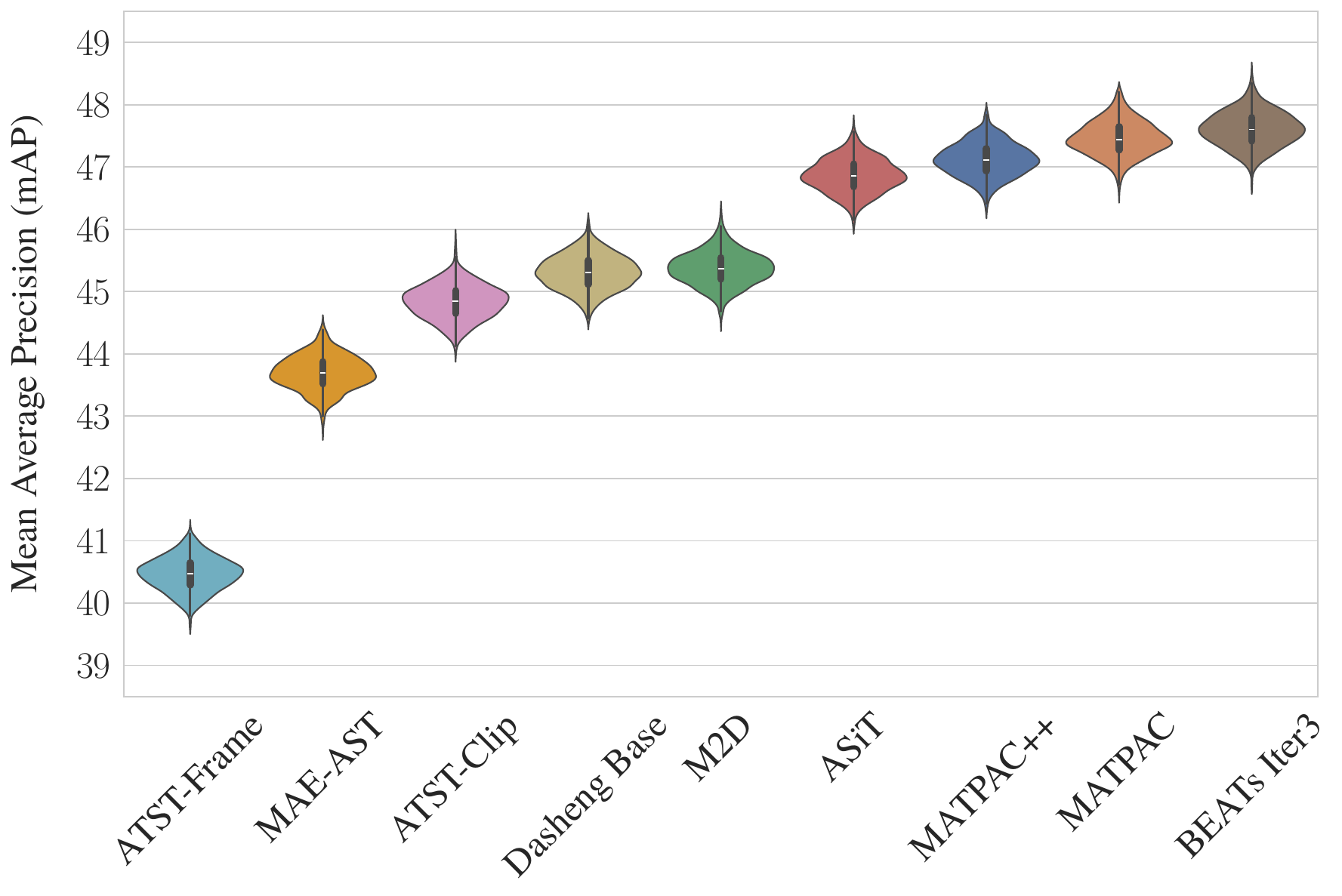}
    \caption{AudioSet bootstrap evaluation for regular fine-tuning.}
    \label{fig:as_bootstrap_ft}
\end{figure}

For both fine-tuning settings, we perform a bootstrapping-based evaluation for all methods considered. This approach involves repeatedly resampling the test set with replacement to estimate the distribution of each model's performance. Beyond providing mean scores, bootstrapping offers valuable insights into the variability and robustness of the results, enabling a more nuanced comparison across methods and a better understanding of how each model responds to variations in the evaluation data. Fig. \ref{fig:as_bootstrap_po_ft} presents the bootstrap evaluation results for the patchout fine-tuning and Fig. \ref{fig:as_bootstrap_ft} for standard fine-tuning settings.

The bootstrap results on the AudioSet test set corroborate the findings of the Wilcoxon signed-rank test: methods with \textit{p}-values between 0.01 and 0.05 exhibit statistically close performances. 

Under the patchout fine-tuning setting (Fig. \ref{fig:as_bootstrap_po_ft}), \acs{met}, MATPAC, and M2D show similar distribution profiles, with \acs{met} consistently achieving the highest scores. This supports the effectiveness of \ac{mlp}-based pretext tasks, while also highlighting the added benefit of combining multi-hypothesis \ac{mcl} with an unsupervised classification objective.

In contrast, the standard fine-tuning setting (Fig. \ref{fig:as_bootstrap_ft}) reveals more pronounced differences in performance distributions, reflected by \textit{p}-values below 0.01 between the best-performing method and the rest. Here, a gap emerges between the top four models (BEATs Iter3, MATPAC, \acs{met}, and ASiT) and the others.
This separation is \secor{particularly evident}{also clear} between \acs{met}, MATPAC, which use both an \ac{mlp} and an unsupervised classification pretext task, and M2D, which relies solely on the \ac{mlp} objective. These results emphasise the complementary strengths of combining multiple pretext tasks in representation learning.

Overall, \acs{met} achieves the highest score on AudioSet across all fine-tuning configurations when fine-tuned with patchout. 

\subsection{Complementary ablations and Results}
\label{app:other_ablations}

\textbf{Impact of the size of the audio segments during pre-training.} To ensure we could perform our ablation studies by pre-training on audio segments of 3 seconds, we compare the results with pre-training on 6 seconds, and we report the results in Table \ref{tab:3vs6}. We observe strongly correlated results, with an average score of 75.5 when using 3-second audio segments compared to an average score of 75.7 when using 6-second audio segments. 

\begin{table}[h!]
    \footnotesize
    \caption{\acs{met} performances when pre-trained with 6-second or 3-second audio segments.}
    \centering
    \resizebox{0.95\textwidth}{!}{
    \begin{tabular}{ l l l l l l l l l l}
             & OpenMIC & NSynth & GTZAN & MTT & FSD50K & ESC50 & US8K & Avg.\\
        Audio length & mAP & Acc(\%) & Acc(\%) & mAP & mAP & Acc(\%) & Acc(\%)\\
        \midrule

        6-second & \textbf{85.6}\pmval{0.1} & 76.8\pmval{0.2} & \textbf{87.6}\pmval{0.0} & 40.8\pmval{0.1} & \textbf{56.1}\pmval{0.1} & \textbf{93.1}\pmval{0.1} & 89.7\pmval{0.3} & \textbf{75.7} \\

        3-second & \textbf{85.6}\pmval{0.0} & \textbf{77.1}\pmval{0.2}$^{\ast\ast}$ & 86.1\pmval{0.3} & \textbf{40.9}\pmval{0.0}$^{\ast}$ & 55.8\pmval{0.0} & 92.8\pmval{0.1} & \textbf{90.3}\pmval{0.1} & 75.5 \\

        \midrule
        
    \multicolumn{10}{l}{$^{\ast}$ $p < 0.05$, $^{\ast\ast}$ $p < 0.01$ compared to best model when CI are overlapping}      
    \end{tabular}
    }
    \label{tab:3vs6}
\end{table}

\begin{figure}[h]
    \centering
    \includegraphics[width=0.8\linewidth]{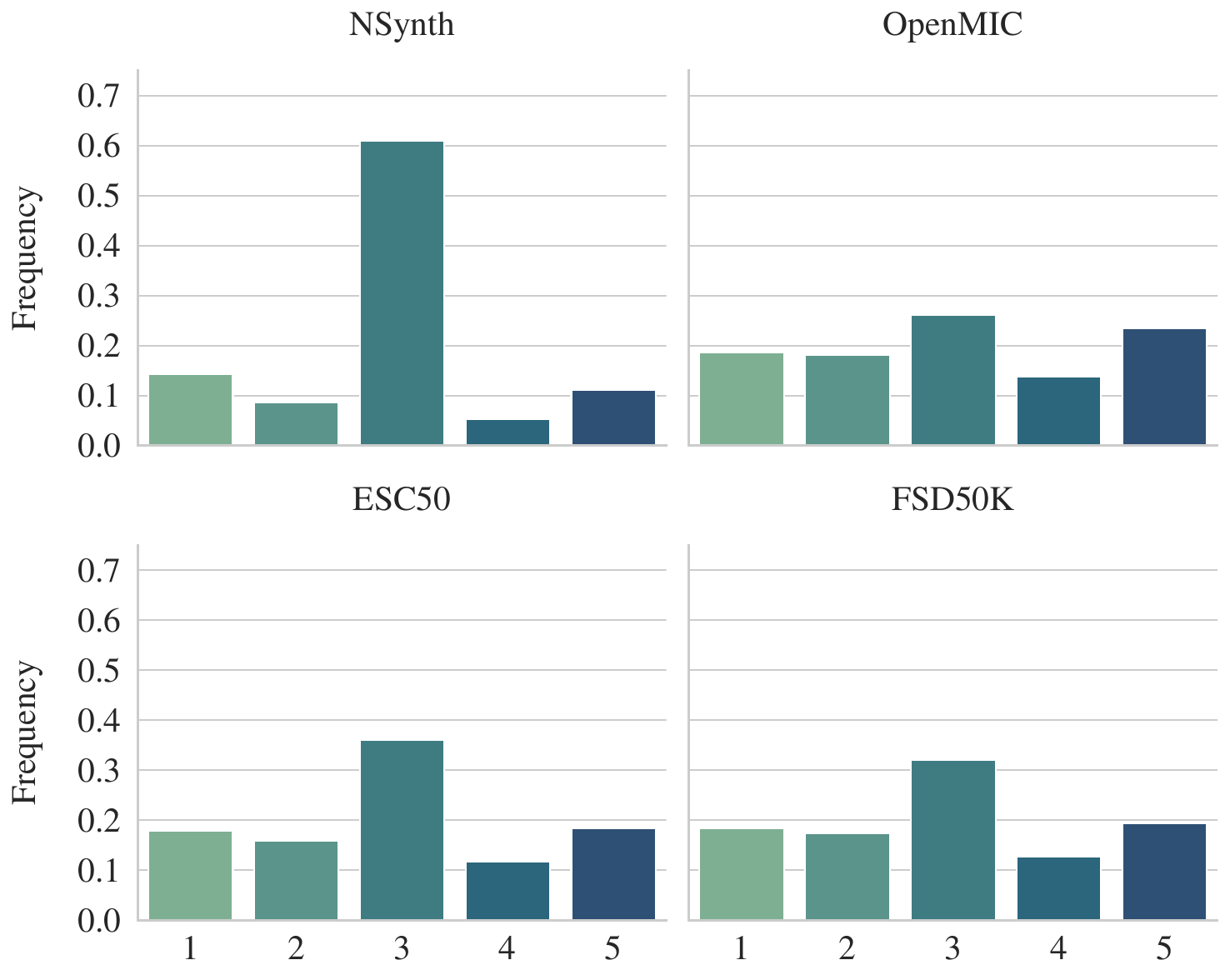}
    \caption{Normalized histogram of the best hypotheses over different downstream datasets. The x-axis is the index of the hypothesis. The y-axis is their normalized selection frequency.}
    \label{fig:head_density}
\end{figure}

\textbf{Hypothesis utilisation frequency.} 
\secmt{Use frequency instead of density everywhere, including Figure}Building on the analysis in Section \ref{sec:ablation}, which showed that each hypothesis specializes in distinct time-frequency patterns, we conduct a follow-up experiment to examine hypothesis usage across downstream tasks. 
We track the best-selected hypothesis for each masked patch and plot their normalized frequency of selection per downstream task, as shown in Fig. \ref{fig:head_density}. 
We select two music tasks (NSynth and OpenMIC) and two environmental tasks (ESC-50 and FSD50K). For each domain, the first downstream task is simple, and the second is complex.
NSynth and ESC-50 are multi-class single-label datasets with isolated sources, while OpenMIC and FSD50K are multi-label and involve polyphonic, real-world recordings. 
This comparison allows us to investigate how hypothesis selection varies with audio complexity and domain.

\begin{figure}[h]
    \centering
    \begin{subfigure}[b]{0.45\linewidth}
        \centering
        \includegraphics[width=\linewidth]{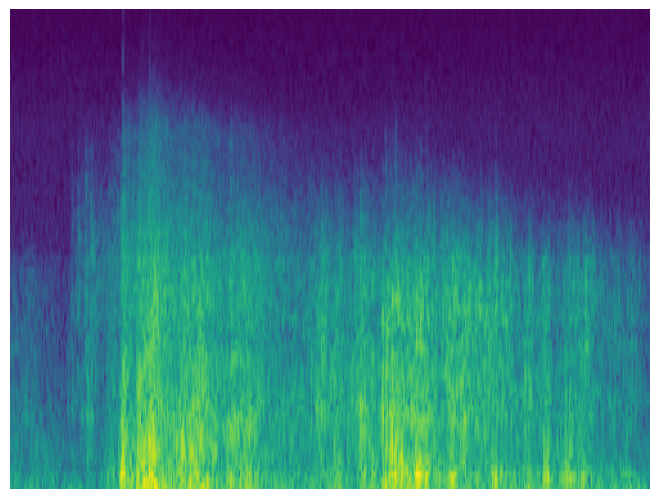}
        \label{fig:sub1}
    \end{subfigure}
    \hfill
    \begin{subfigure}[b]{0.45\linewidth}
        \centering
        \includegraphics[width=\linewidth]{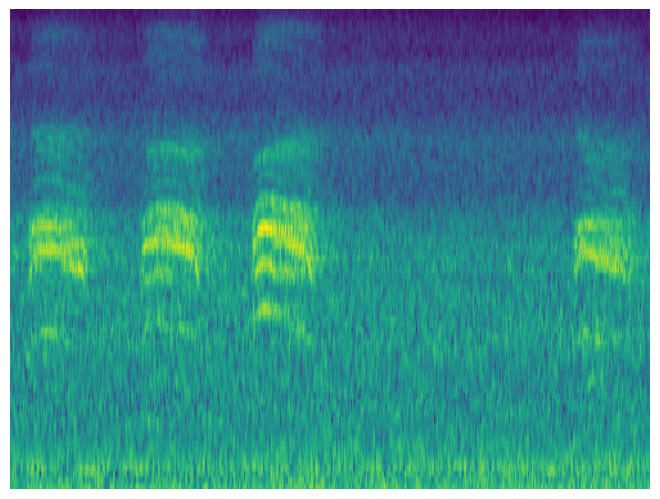}        
        \label{fig:sub2}
    \end{subfigure}
    \vskip\baselineskip
    \begin{subfigure}[b]{0.45\linewidth}
        \centering
        \includegraphics[width=\linewidth]{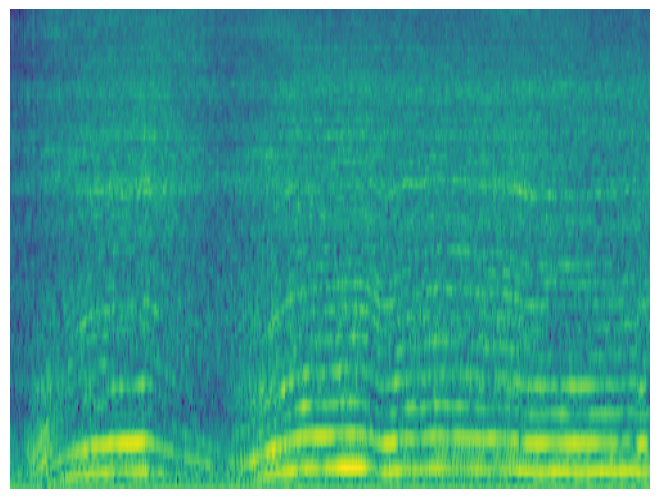}        
        \label{fig:sub3}
    \end{subfigure}
    \hfill
    \begin{subfigure}[b]{0.45\linewidth}
        \centering
        \includegraphics[width=\linewidth]{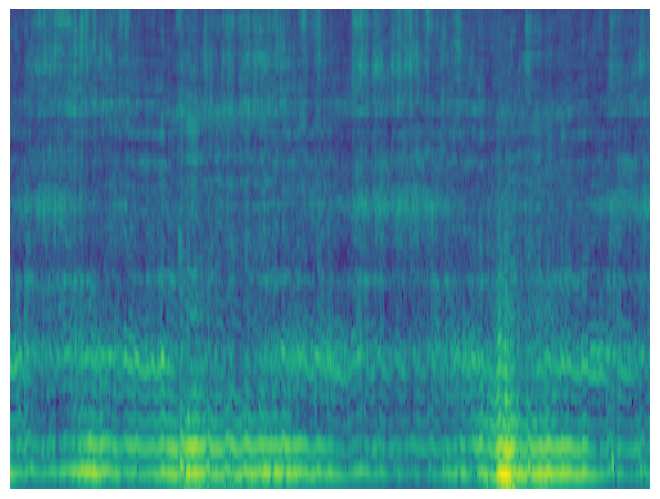}        
        \label{fig:sub4}
    \end{subfigure}
    \caption{Log-scale Mel Spectrogram for different audio samples of ESC-50. x-axis is the time and y-axis the frequencies.}
    \label{fig:esc50}
\end{figure}

\begin{figure}[h]
    \centering
    \begin{subfigure}[b]{0.45\linewidth}
        \centering
        \includegraphics[width=\linewidth]{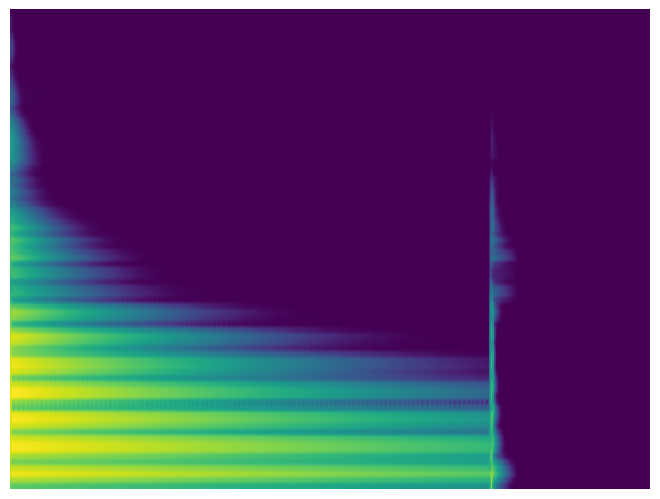}
        \label{fig:sub1}
    \end{subfigure}
    \hfill
    \begin{subfigure}[b]{0.45\linewidth}
        \centering
        \includegraphics[width=\linewidth]{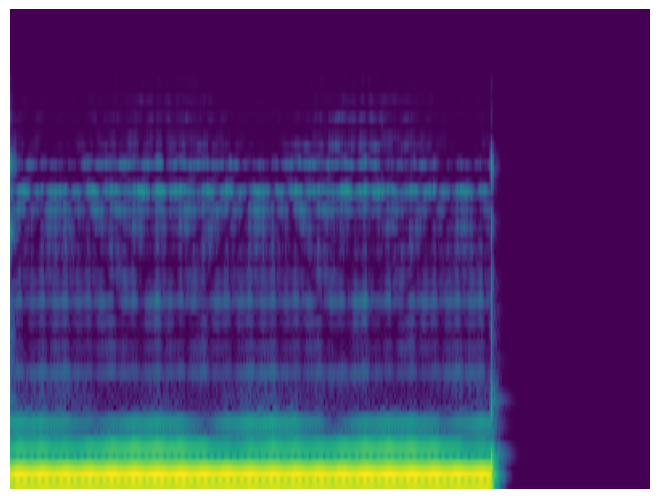}        
        \label{fig:sub2}
    \end{subfigure}
    \vskip\baselineskip
    \begin{subfigure}[b]{0.45\linewidth}
        \centering
        \includegraphics[width=\linewidth]{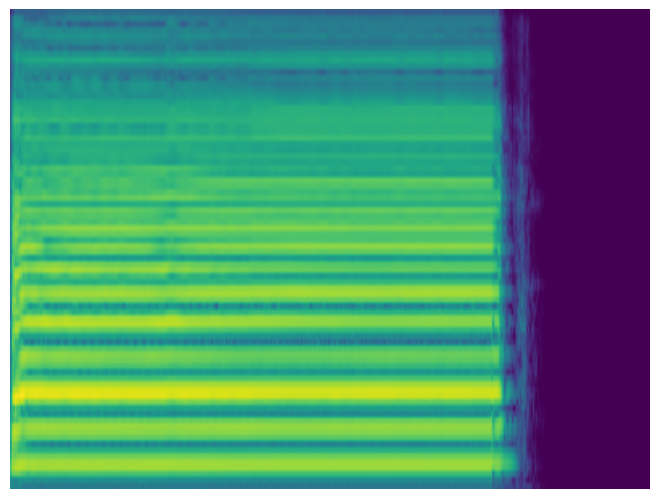}        
        \label{fig:sub3}
    \end{subfigure}
    \hfill
    \begin{subfigure}[b]{0.45\linewidth}
        \centering
        \includegraphics[width=\linewidth]{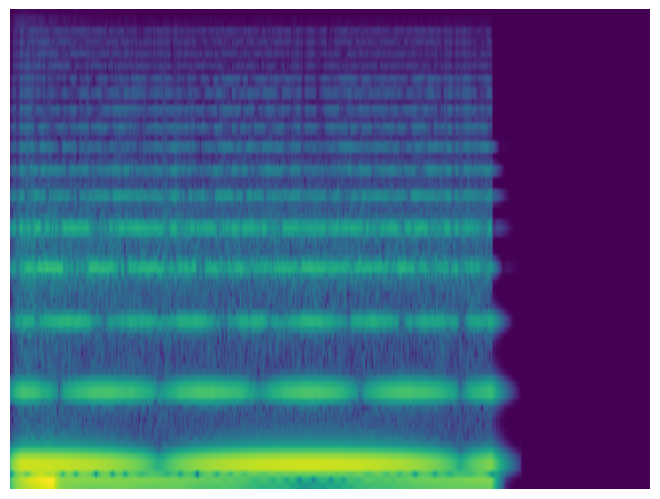}        
        \label{fig:sub4}
    \end{subfigure}
    \caption{Log-scale Mel Spectrogram for different audio samples of NSynth. x-axis is the time and y-axis the frequencies.}
    \label{fig:nsynth}
\end{figure}

\begin{figure}[h]
    \centering
    \begin{subfigure}[b]{0.45\linewidth}
        \centering
        \includegraphics[width=\linewidth]{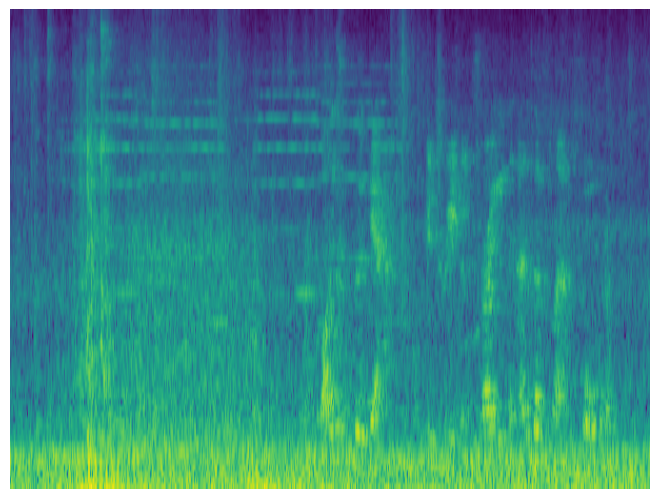}
        \label{fig:sub1}
    \end{subfigure}
    \hfill
    \begin{subfigure}[b]{0.45\linewidth}
        \centering
        \includegraphics[width=\linewidth]{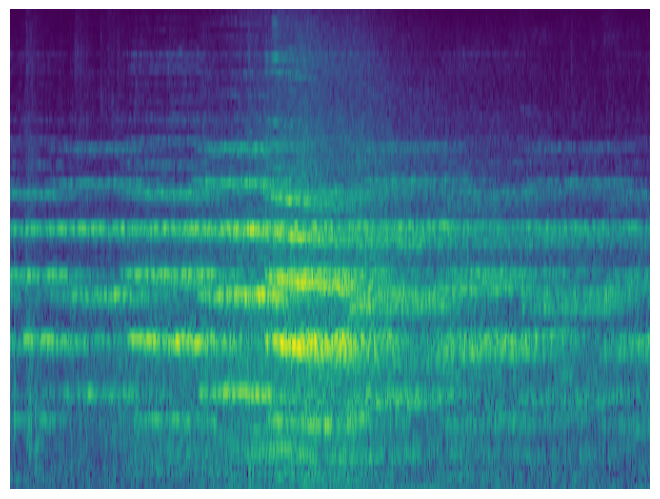}        
        \label{fig:sub2}
    \end{subfigure}
    \vskip\baselineskip
    \begin{subfigure}[b]{0.45\linewidth}
        \centering
        \includegraphics[width=\linewidth]{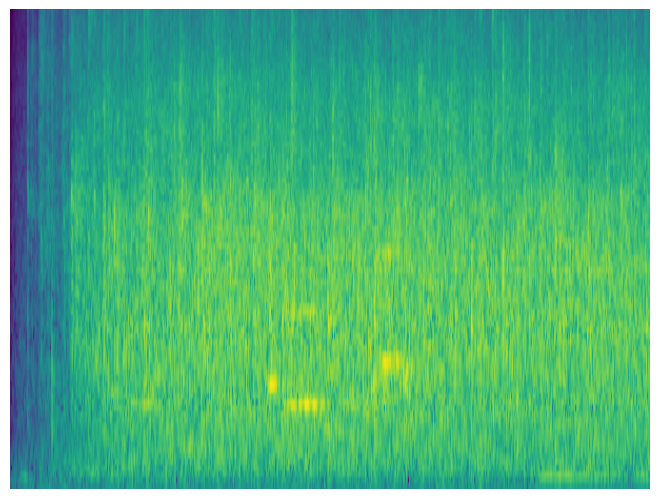}        
        \label{fig:sub3}
    \end{subfigure}
    \hfill
    \begin{subfigure}[b]{0.45\linewidth}
        \centering
        \includegraphics[width=\linewidth]{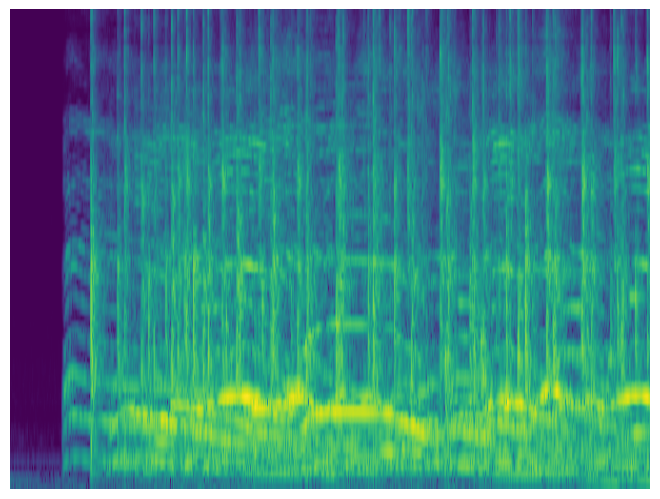}        
        \label{fig:sub4}
    \end{subfigure}
    \caption{Log-scale Mel Spectrogram for different audio samples of FSD50K. x-axis is the time and y-axis the frequencies.}
    \label{fig:fsd50k}
\end{figure}

\begin{figure}[h]
    \centering
    \begin{subfigure}[b]{0.45\linewidth}
        \centering
        \includegraphics[width=\linewidth]{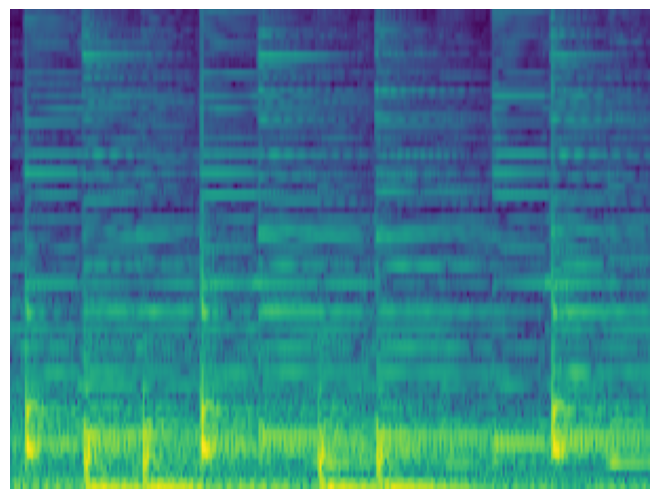}
        \label{fig:sub1}
    \end{subfigure}
    \hfill
    \begin{subfigure}[b]{0.45\linewidth}
        \centering
        \includegraphics[width=\linewidth]{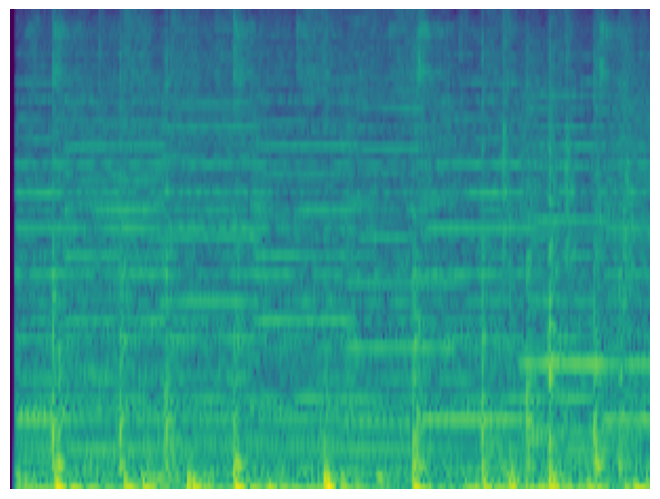}        
        \label{fig:sub2}
    \end{subfigure}
    \vskip\baselineskip
    \begin{subfigure}[b]{0.45\linewidth}
        \centering
        \includegraphics[width=\linewidth]{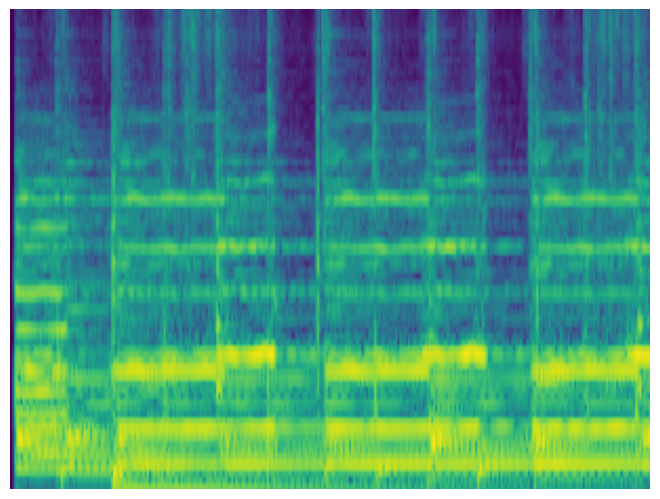}        
        \label{fig:sub3}
    \end{subfigure}
    \hfill
    \begin{subfigure}[b]{0.45\linewidth}
        \centering
        \includegraphics[width=\linewidth]{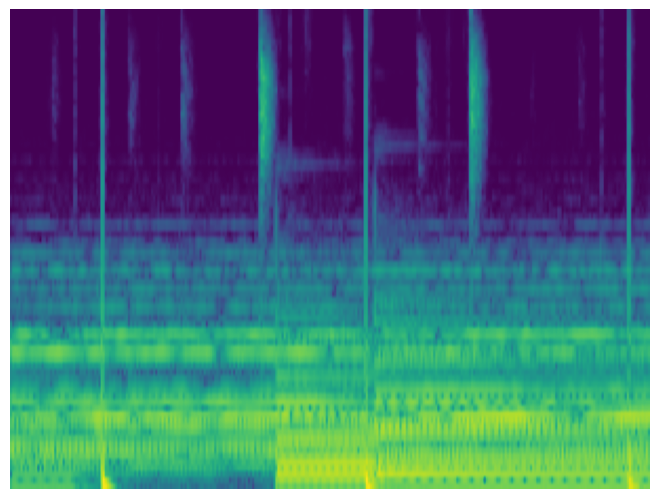}        
        \label{fig:sub4}
    \end{subfigure}
    \caption{Log-scale Mel Spectrogram for different audio samples of OpenMIC. x-axis is the time and y-axis the frequencies.}
    \label{fig:openmic}
\end{figure}

Overall, we observe a consistent pattern across tasks: hypothesis 3 is often selected. 
One explanation is that the corresponding patches mostly represent uniformly distributed noise, as hypothesis 3 is specialised to predict patches with predominantly constant energy (see Fig. \ref{fig:prototypes}).
In simpler datasets like NSynth and ESC-50, where only one source is active and the time-frequency content is either limited or with uniform noise level (see Fig. \ref{fig:esc50} and Fig. \ref{fig:nsynth}), hypothesis 3 is activated more often than in complex, multi-source tasks. 
Additionally, when comparing OpenMIC and FSD50K, hypothesis 5 is more selected in OpenMIC. 
This likely reflects its specialisation in predicting patches with energy distributed along the frequency axis (see Fig. \ref{fig:prototypes}).
Indeed, music contains structured harmonic and percussive content, compared to the less structured nature of environmental sounds, and therefore has more patterns where the energy is distributed along the frequencies (see Fig. \ref{fig:fsd50k} and Fig. \ref{fig:openmic}). 

These results further underscore the effectiveness of incorporating multiple hypotheses via \ac{mcl} in the \ac{mlp} pretext task, as it both addresses prediction ambiguity and encourages specialisation among hypotheses.

\section{Evaluation Details}
\label{app:expdetails}

\subsection{Pre-training Datasets}
\label{app:pre_datasets}

In Table \ref{tab:pretrain} we report the information related to the dataset used for pre-training our method. We list the number of samples that are longer than 6 seconds, \ie the ones we used for pre-training. In the ``Pre-training'' column, we mention for which pre-training the datasets were used, either the general audio pre-training or the music pre-training. 

\begin{table}[htb!]
    \caption{Pre-training datasets. AudioSet is also used as a fine-tuning dataset.}
    \centering    
    \resizebox{1\columnwidth}{!}{
    \begin{tabular}{c| c c c c c c c}
        Dataset & Pre-training & \#Samples & \#Classes & Content & Accessibility \\
        \midrule
        AudioSet \cite{gemmeke2017audioset} & General Audio & 2012615 & 527 & General Audio from Youtube Videos. & Public     \\ 

        \addlinespace[0.07cm] \hdashline \addlinespace[0.11cm]
        
        MillionSong \cite{Mahieux11msd} & Music & 904130 & - & Western Commercial Music & Public     \\ 
        WebRadio & Music & 180646 & - & Songs from 16 thematic web radios & Private \\
        
    \end{tabular}
    }
    \label{tab:pretrain}
\end{table}

\subsection{Model Parameters}
\label{app:model_params}

We use the same hyperparameters as MATPAC \cite{quelennec2025matpac} and M2D \cite{niizumi23m2d} for the encoders, decoders, classification heads, and training parameters.
However, since we incorporate the multi-hypothesis \ac{mcl} mechanism in its annealed form, it introduces an additional hyperparameter: $\tau_{mcl}$. 
In all experiments, we initialize $\tau_{mcl}$ to 1 and apply an exponential decay with $\eta = 0.99997$, i.e., $\tau_{mcl} \xleftarrow{} \eta\tau_{mcl}$. As detailed in Section~\ref{sec:method_mlp}, this annealing schedule enables broad hypothesis exploration at the beginning of the training, before gradually focusing on the best hypothesis.

All experiments were conducted using 4 NVIDIA H100 GPUs. Pre-training on 6-second audio segments required approximately 30 hours, while training on 3-second segments reduced the time to 18 hours.

\subsection{Downstream Tasks}
\label{app:ds_tasks}

Table \ref{tab:ds_datasets} provides details of each downstream task. MCSL denotes Multi-Class Single Label downstream task, and ML denotes Multi-Label downstream task. 
We also specify the type of split used for each dataset, with TVT standing for Train/Validation/Test. The column ES Crit. reports the early stopping criterion used for each task.

\begin{table}[h]
    \caption{Evaluation Downstream tasks.}
    \centering    
    \resizebox{0.85\columnwidth}{!}{
    \begin{tabular}{c| c c c c c c c}
        Dataset & \#Samples & \#Classes & Type & Split & Metric & ES Crit. & Accessibility \\
        \midrule
        OpenMIC \cite{Humphrey18openmic} & 20,000 & 20 & ML & TVT  & mAP & val mAP & Public\\
        NSynth \cite{EngelRRDNES17nsynth} & 305,979 & 11 & MCSL & TVT & Acc. & val Acc. & Public\\
        GTZAN \cite{TzanetakisC02gtzan} & 930 & 10 & MCSL & TVT & Acc. & val loss & Public\\
        MTT \cite{magnatag} & 25,863 & 50 & ML & TVT & mAP & val loss & Public\\
        FSD50K \cite{FonsecaFPFS22fsd50k} & 51,197 & 200 & ML & TVT & mAP & val mAP  & Public\\
        ESC-50 \cite{Piczak15esc50} & 2,000 & 50 & MCSL & 5-Fold & Acc. & val loss & Public\\
        US8K \cite{SalamonJB14us8k} & 8,732 & 10 & MCSL & 10-Fold & Acc. & val Acc. & Public\\
        MTG Instrument \cite{bogdanov2019mtg} & 24,976 & 40 & ML & TVT & mAP & val mAP & Public\\
        MTG Mood \cite{bogdanov2019mtg} & 17,982 & 56 & ML & TVT & mAP & val mAP  & Public\\
        MTG Genre \cite{bogdanov2019mtg} & 55,094 & 87 & ML & TVT & mAP & val mAP & Public\\
        MTG Top50 \cite{bogdanov2019mtg} & 54,380 & 50 & ML & TVT & mAP & val mAP  & Public\\ 
    \end{tabular}
    }
    \label{tab:ds_datasets}
\end{table}

\subsection{Downstream Evaluation Protocol}
\label{app:ds_protocol}

The downstream evaluation protocol follows the one proposed in \cite{quelennec2025matpac}.
We train a linear classifier composed of a single fully connected layer that maps the model's encoder latent representation to the task-specific classes. 
All tasks use a batch size of 128 and a learning rate of $1\mathrm{e}{-4}$ with the Adam optimiser. 
For \acs{met}, we pre-trained the model using either 6-second or 3-second audio segments.
If the input audio exceeds this duration in downstream tasks, we segment it into non-overlapping chunks (of  either 6 seconds or 3 seconds).
We then apply average pooling over time to produce a single embedding per clip.
Our evaluation protocol differs from that of \cite{quelennec2025matpac} by employing dataset-specific early stopping criteria. 
This criterion depends on the downstream task as described in Table \ref{tab:ds_datasets}.
We either used the best validation loss, or validation accuracy, or validation mAP.

\subsection{Fine-tuning Protocol}
\label{app:ft_protocol}

When fine-tuning the encoders, we use the code and the parameters proposed by \cite{niizumi24m2dx}.
We report the main parameters in Table \ref{tab:ft-parameters}. 
For fine-tuning, we use the AudioSet \cite{gemmeke2017audioset} dataset with the labels (AudioSet is also used for pre-training, but without the labels). 

\begin{table}[h]
\caption{Encoder Finetuning settings.}
\label{tab:ft-parameters}
\centering
\resizebox{0.4\columnwidth}{!}{%
\begin{tabular}{llllll}
\toprule
Parameter & AS2M  \\
\midrule
Learning rate & 2.0 \\
Batch size & 64 \\
Optimizer & LARS \\
Mixup ratio & 0.5 \\
SpecAugment$^\sharp$\cite{specaugment} & 30/192 \\
Training epochs (total) & 70 \\
Training epochs (warm-up) & 15\\
Patchout masking ratio & 0.5 \\
\bottomrule
\addlinespace[0.05cm]
\multicolumn{6}{l}{$^{\sharp}$ The frequency/time masking parameters.}\\
\end{tabular}
}
\end{table}

All fine-tuning experiments ran on a single H100 with a time limit of 100 hours.
Every model was able to converge under these settings except Dasheng 0.6B \cite{dinkel2024dasheng}.
For this reason, we did not report the results of Dasheng 0.6B in Fig. \ref{fig:as_bootstrap_ft}.

\section{Detailed Related Works}
\label{app:relatedworks}

\textbf{Masked latent prediction pretext tasks.}
Methods that operate in the latent space typically use a teacher-student architecture and aim to maximise the agreement between their representations. 
Some approaches, such as ATST-CLIP \cite{LiSL24atst} and BYOL-A \cite{Niizumi23byola}, enforce consistency between two views of the same audio.
These views are obtained through temporal cropping or data augmentation. 
A more effective family of methods is based on Masked Latent Prediction (\ac{mlp}), where the student encodes only a visible part of the input, and a predictor aligns the student's output with the teacher's latent representation of the input. 
This idea was first introduced by Data2Vec and has been extended by models such as ATST-Frame \cite{LiSL24atst}, which applies augmentations to the student input.
I-JEPA \cite{assran23ijepa} and M2D \cite{niizumi23m2d} improved this design by encoding only the visible patches with the student and the masked patches with the teacher, improving both performances and efficiency.
In our work, we adopt the \ac{mlp} paradigm but extend it to address its inherent ambiguity by introducing a multi-hypothesis prediction framework based on \ac{mcl}.

\textbf{Unsupervised classification pretext tasks.}
Classification-based methods use ``pseudo'' labels.
BEATs \cite{chen23beats}, which operates on general audio, uses a learned tokenizer for those.
In HuBERT \cite{Hsu21HUBERT}, which operates on speech, a k-means clustering on audio features is used for those.
Both methods need to be trained iteratively to achieve optimal performance. 
In computer vision, iBOT \cite{Zhou22ibot} extends DINO \cite{Caron21dino} with a masked image modelling objective and a classification pretext task at the image patch level. 
In contrast, \cite{Oquaba24dinov2} combines DINO and iBOT losses to have local-level and global-level classification, while enabling scaling capabilities. 
Inspired by these advances, our method incorporates an unsupervised classification objective into the audio domain. 
This objective is designed to operate jointly with \ac{mlp} and is enhanced by a multi-hypothesis learning framework.

\textbf{Multiple pretext tasks.}
To learn robust representations, \ac{ssl} methods often combine multiple pretext tasks to leverage their complementary strengths. For example, ASiT \cite{Ahmed24Asit} and MAE-AST \cite{Baade22maeast}, jointly apply a reconstruction-based objective inspired by Masked Auto-Encoding (MAE) alongside a contrastive classification loss.
The classification task encourages each token from the student encoder to align with its target while remaining distinct from others. 
These classification objectives typically operate at the patch level and assign instance-specific class. 
MATPAC \cite{quelennec2025matpac} introduces a cascaded approach, first applying \ac{mlp}, then performing an unsupervised classification pretext task.
The student encoder, followed by the predictor, and the teacher encoder produce predicted and target representations first.
Then they are separately projected by a student and teacher classification head into probability distributions.
Finally, those are matched through a classification loss.
In this case, pseudo-classes are derived from projected latent vectors, similar to DINO but without requiring multiple target views.
Building on this framework, our method enhances the \ac{mlp} stage by incorporating \ac{mcl}, enabling the model to generate multiple plausible hypotheses. 

\textbf{Multiple Choice Learning.}
\ac{mcl} approaches address task ambiguity by generating multiple candidate solutions (hypotheses).
Usually, only the solution with the lowest loss is used for optimisation. 
This strategy is commonly referred to as greedy hypothesis selection \cite{guzman12mcl, Lee16mcl, rupprecht17mcl, garcia21mcl, letzelter23mcl}. 
However, this greedy selection can lead to hypothesis collapse.
To mitigate this issue, recent advances such as annealed \ac{mcl} \cite{perera24mcl}, encourage broader exploration during training. 
Following these findings, our work integrates annealed \ac{mcl} into the \ac{mlp} framework. 
This allows to model ambiguity and explicitly promote diverse, task-relevant representations.

\end{document}